\documentclass[final]{raa}           

\usepackage{graphicx,times}
\usepackage{natbib}
\usepackage{amssymb,amsmath}
\usepackage{caption}
\bibpunct{(}{)}{;}{a}{}{,}
\def\gtorder{\mathrel{\raise.3ex\hbox{$>$}\mkern-14mu
             \lower0.6ex\hbox{$\sim$}}}
\def\ltorder{\mathrel{\raise.3ex\hbox{$<$}\mkern-14mu
             \lower0.6ex\hbox{$\sim$}}}

\def\au{{\rm AU}}

\def\mas{{\rm mas}}

\def\muas{\mu{\rm as}}

\def\rel{{\rm rel}}

\def\Le{{\rm L}}
\def\E{{\rm E}}

\def\csst{{\it CSST}}
\def\rst{{\it Roman}}

\begin{document}

   \title{Detecting Exomoons in Free-Floating-Planet Events from Space-based Microlensing Surveys}

 \volnopage{ {\bf 20XX} Vol.\ {\bf X} No. {\bf XX}, 000--000}
   \setcounter{page}{1}

   \author{Haozhu Fu %(付皓竹) %% Put your Chinese name in "( )" if you like. Note to open line 11 "\usepackage[UTF8]{ctex}"
   \inst{1,2}, Subo Dong\inst{1,2,3,4}
   }

   \institute{ Department of Astronomy, Peking University, 5 Yiheyuan Road, Haidian District, Beijing 100871, 
China\\
        \and
             Kavli Institute of Astronomy and Astrophysics, Peking University, 5 Yiheyuan Road, Haidian District, Beijing 100871, China \\
        \and 
        National Astronomical Observatories, Chinese Academy of Science, 20A Datun Road, Chaoyang District, Beijing 100101, China\\
        \and
        Corresponding Author; {\it dongsubo@pku.edu.cn}\\
\vs \no
   {\small Received 20XX Month Day; accepted 20XX Month Day}
}
\abstract{When a planet is ejected from its star-planet system due to dynamical interactions, its satellite may remain gravitationally bound to the planet. The {\it Chinese Space Station Telescope} (\csst) will be capable of detecting a large number of low-mass free-floating planet events (FFPs) from a bulge microlensing survey. We assess the feasibility of detecting satellites (a.k.a., exomoons) orbiting FFPs by simulating \csst\ light curves and calculating the detection efficiency as a function of satellite-to-planet mass ratios $(q)$ and projected separations $(s)$ in units of the Einstein radius. For a Neptune-class FFP in the Galactic disk with a Sun-like star as the microlensed source, \csst\ can detect Earth-mass satellites over a decade of separations ($\sim 0.01$--$0.1$\,\au) and has sensitivity down to Moon-mass satellites ($q\sim10^{-3}$) at $s\sim1$. \csst\ also has some sensitivity to detect Moon-mass satellites at $s\sim2$ ($\sim 0.02$\,\au) orbiting an Earth-mass FFP in the disk. \csst\ has substantially reduced sensitivity for detecting satellites when the source star is an M dwarf, compared to a Sun-like source. We also calculate the satellite detection efficiency for the dedicated microlensing survey of the {\it Roman Space Telescope} (\rst), which demonstrates greater sensitivity  than \csst, particularly for M-dwarf sources. Notably, some of the Neptune-Earth systems detectable by \csst\ and {\it Roman} may exhibit significant tidal heating.
}

   \authorrunning{Hao-Zhu Fu and Subo Dong}  
      \titlerunning{Detecting Exomoons in FFP Microlensing Events}     \maketitle

\section{Introduction}   \label{sec:intro}

A planet gravitationally unbound to any host stars can be detected in a short-duration ($<1$\,d) microlensing event. In the event, the planet acts as a ``lens'' bending the light from a background star (``source'') in the observer's sightline. More than a dozen such free-floating-planet candidates (hereafter FFPs) have been found (for recent reviews, see \citealt{ZhuDong21, ulensplanetreview}). An FFP event can also be due to a planetary lens on a wide-separation ($\gtrsim20$\,\au) orbit from a host star, which can be distinguished from an unbound planet by follow-up adaptive-optics observations with instruments such as ELT-MICADO \citep{MICADO} a few years after the event.  Statistical analyses on three independent samples from OGLE, KMTNet, and MOA surveys suggest that low-mass (Earth-mass to Neptune-class) FFPs are likely common in the Galaxy, possibly a few times more numerous than stars \citep{Mroz17, Gould22, Sumi23}.  

A class of mechanisms to generate unbound planets is dynamical ejection from planetary systems via planet-planet scatterings \citep[e.g.,][]{Rasio96,Weidenschilling96,Juric08, Chatterjee08}. The ejection likely occurs from a perturber in the outer planetary system ($a>1$\,\au) with the Safronov number larger than unity, that is, the escape velocity at the perturber's surface being greater than the orbital escape velocity. Previous works \citep{Rabago19,Debes07,Hong18} suggest that, when a planet is ejected, its satellite has a reasonable probability of remaining gravitationally bound to the planet. Free-floating planet-satellite systems are also studied for habitability \citep{Reynolds87,Scharf06} in that they might be able to maintain the existence of liquid water due to tidal heating \citep{Reynolds87,Scharf06}.
    
To date, there have been a handful of satellite candidates (i.e., exomoons) around bound planets identified by microlensing or transits but with no definitive detection \citep[see, e.g.,][]{Bennett14, Teachey18, Kreidberg19}. Because the finite-source effects \citep{Gould94, Nemiroff94, WittMao94} could significantly reduce the amplitude of a satellite's signal, a space-based microlensing survey offers the most promising opportunity for detecting satellites via microlensing \citep{BennettRhie02, Han02, Liebig10}.

In this paper, we focus on studying the prospects of detecting satellites orbiting FFPs in a bulge microlensing survey using the {\it Chinese Space Station Telescope} (\csst)\footnote{{https://nadc.china-vo.org/csst-bp/article/20230707113736}}, which is a planned 2-m space telescope in a low-Earth orbit ($\approx1.5$\,hr period).  \csst\ will be equipped with a wide-field ($\sim1.1$\,deg$^2$) survey camera designed to take diffraction-limited images at optical wavelengths. We also conduct simulations for {\it the Nancy Grace Roman Space Telescope} (a.k.a., {\it WFIRST}; hereafter {\it Roman}), whose Galactic Bulge Time Domain Survey aims to discover bound planets and FFPs with microlensing \citep{Penny19, Johnson20, YeeGould23}. Recently, \citet{Sajadian23} studied \rst's detection efficiency of free-floating planet-satellite systems, and we compare our results with theirs.

\section{Simulating Microlensing Observations}
\label{sec:Obs}

The Einstein radius sets the basic scale in microlensing, and its angular size is,
\begin{equation}
\theta_\E=\sqrt{\kappa M_\Le\pi_\rel},
\end{equation}\label{e2}
where 
\begin{equation}
\pi_\rel={\au}({\frac{1}{D_\Le}}-\frac{1}{D_{\rm S}}),
\end{equation}
is the relative trigonometric parallax between the lens at distance $D_\Le$ and the source at $D_{\rm S}$, $M_\Le$ is the lens mass and $\kappa=4\pi G/(c^2 \au )=8.144\,\mas\,M_\odot^{-1}$ is a constant. The physical Einstein radius is $R_\E=D_\Le \theta_\E$.

The ultra-short microlensing events with measured $\theta_\E$ provide compelling evidence for the existence of FFPs \citep[e.g.,][]{Mroz2018}. They collectively have $\theta_\E\lesssim 9\muas$, below an empirical gap in the $\theta_\E$ distribution between $\sim 9\muas$ and $\sim 25\muas$ (the ``Einstein desert''), suggesting that they belong to a planetary population separated from brown dwarfs and low-mass stars \citep{Ryu21, Gould22}. The $\theta_\E$ measurements are made for finite-source-point-lens (FSPL) events, during which the lens transits the source with angular radius $\theta_*$ being larger than or comparable to $\theta_\E$. From fitting the light curve of an FSPL event, the scaled source size $\rho=\theta_*/\theta_\E$ can be directly extracted, and then $\theta_\E$ is estimated by using $\theta_*$ measured via the source's color and apparent magnitude \citep{Yoo04a}. 

In an FSPL event, the magnification $A(t)$ is a function of  $(t_0,u_0,t_E,\rho)$, where $t_\E=\theta_E/\mu_\rel$ is the time taken by the source at a relative proper motion $\mu_\rel$ with respect to the lens to cross $\theta_\E$, $u_0$ is the impact parameter and $t_0$ is the time of the peak. Introducing a satellite into the FFP lens system requires three additional binary-lens parameters $(s, q, \alpha)$, where $q={m_{\rm satellite}}/{m_{\rm planet}}$ is the satellite-planet mass ratio, $s$ is the angular distance between the planet and the satellite scaled by $\theta_{\rm E}$, and $\alpha$ is the angle between the trajectory of the source and the satellite-planet vector.

We adopt the approach of \citet{Yan22} to simulate the surveys. We estimate \csst\ photometric uncertainties using the Exposure Time Calculator\footnote{{https://nadc.china-vo.org/csst-bp/etc-ms/etc.jsp}} for an exposure time of 60s in $i$-band and a systematic noise floor of 0.001 mag. The duty cycle is 40\%, and for each orbit, there are 8 observations (i.e., $\sim$5-min cadence). We adopt a 15-min cadence for \rst\, and assume the photometric performance in $W149$ according to \citet{Penny19}.

We simulate microlensing events with a set of representative parameters for the source and lens. Following \citet{Yan22}, we consider two types of sources: early M-dwarfs (M0V) and Sun-like dwarfs (G2V) in the Galactic bulge ($D_S=8.2\,\rm kpc$), adopting the source sizes of $\theta_*=0.28\,{\rm \mu as}$ and $0.56\,{\rm \mu as}$, respectively. We first estimate the Johnson-Cousin $I$- and $H$-band magnitudes  based on \citet{2013ApJS..208....9P}, and we apply extinction corrections by adopting $A_I=1.5$ and $E(I-H)=1$ \citep{Gonzalez12,Nataf13}. Then we convert the Johnson-Cousin magnitudes in the Vega system to the AB system using results from \citet{2007AJ....133..734B}, deriving that the M0V (G2V) source has $m_i \approx 23\,(20)$\,mag and $m_{W149} \approx 21.7\,(19.6)$\,mag, corresponding to uncertainties in baseline magnitudes of $\approx0.05\,(0.01)$\,mag and $\approx0.014\,(0.005)$\,mag, respectively. 

The lens is placed either in the bulge or the disk. For a bulge lens in our simulations, the relative parallax is $\pi_{\rm rel}=0.02$\,mas ($D_\Le\approx7\,{\rm kpc}$) with a relative proper motion of $\mu_{\rm rel}=4\,{\rm mas/yr}$. For a disk lens, $\pi_{\rm rel}=0.12$\,mas ($D_\Le\approx4\,{\rm kpc}$) and $\mu_{\rm rel}=7\,{\rm mas/yr}$. The FFP lens is assumed to have a mass of either an Earth-mass planet ($M_{\rm L}=3\times10^{-6}\,M_\odot$) or a Neptune-class (i.e., using the averaged mass of Neptune and Uranus) planet ($M_{\rm L}=4.8\times10^{-5}\,M_\odot$). The adopted planet properties are listed in Table~\ref{lens}, while the parameters adopted in our simulations are listed in Table~\ref{parameters}.

\begin{table}[h]
    \centering
    \begin{minipage}{0.4\textwidth}
        \centering
        \caption{\label{source}Adopted Source Properties }
        \begin{tabular}{llll}
            \hline
            \hline
            source type & $\theta^*/\mu$as & $i$ mag & $W149$ mag\\
            G2V         & 0.56             & 20            & 19.6\\
            M0V         & 0.28             & 23            & 21.7\\
            \hline
        \end{tabular}
    \end{minipage}
    \hfill
    \begin{minipage}{0.55\textwidth}
        \centering
        \caption{\label{lens}Adopted Planet Masses}
        \begin{tabular}{llll}
            \hline
            \hline
            planet type   & mass$/M_\odot$     & mass$/M_\oplus$ & $\log({{\rm mass}/m_{\rm Moon}})$ \\
            Earth-mass    & $3\times10^{-6}$   & 1               & $1.91$\\
            Neptune-class & $4.8\times10^{-5}$ & 16              & $3.11$\\
            \hline
        \end{tabular}
    \end{minipage}
\end{table}

\begin{table}[h]
\begin{center}
\caption{\label{parameters}Microlensing Parameters Adopted in the Simulations}
\begin{tabular}{llllll}
\hline
\hline
planet type & $\pi_{\rm rel}/$mas (planet location) & $\theta_\E/{\rm \mu as}$ & $R_\E/\au$ & $t_\E/{\rm day}$ & $\rho\,{\rm (source\, type)}$        \\ \hline
Earth-mass       & 0.12 (disk)    &  1.70 & 0.007  & 0.090  & 0.33 (G2V) \& 0.165 (M0V)  \\
Neptune-class  & 0.12 (disk)     &  6.85 & 0.03   & 0.357  &  0.08 (G2V) \& 0.04 (M0V) \\ 
Earth-mass      & 0.02 (bulge)  &  0.70 & 0.005  & 0.064  & 0.80 (G2V) \& 0.40 (M0V)  \\
Neptune-class  & 0.02 (bulge)  & 2.80  & 0.02    & 0.256  & 0.20 (G2V) \& 0.10 (M0V)  \\
\hline
\end{tabular}
\end{center}
\end{table}
\subsection{Detection efficiency}

We follow the commonly used method \citep{Rhie00, Dong06} to estimate the detection efficiency of satellites. Mock binary-lens (i.e., planet-satellite) single-source (2L1S) light curves are simulated using the \texttt{VBBinaryLensing} code \citep{Bozza18}. Then the mock data are fitted to single-lens single-source (1L1S) models with free parameters $(t_0, u_0, t_\E, \rho)$ using Markov chain Monte Carlo (MCMC). The magnifications for 1L1S models with finite-source effects (i.e., FSPL models) are calculated with the map-making algorithm \citep{Dong06, Dong09}. A satellite is regarded as being detected if the $\chi^2$ difference between the best-fit 1L1S model and the input 2L1S model exceeds the detection threshold ($\Delta{\chi^2} \geq 100$). For each set of $(s, q)$, we evaluate 180 uniformly distributed $\alpha$ values within the range $[0, 2\pi)$. The detection efficiency for a given parameter set is defined as the fraction of $\alpha$ values for which the satellite is detected. Then, we calculate the detection-efficiency distribution in the $s$-$q$ plane, covering the logarithmic ranges $[-1.0,1.0]$ for $\log{s}$ and $[-4.0,-1.0]$ for $\log{q}$, with step sizes of 0.033 dex and 0.1 dex, respectively. We analyze mock data with $u_0=0$ by default and evaluate how  non-zero $u_0$ impacts the detection efficiency. 

\section{Simulation Results}
\label{sec:results}

In this section, we present the results of our simulations. In contrast to the ground-based surveys, which are most sensitive to FFPs with bright giant sources, space-based surveys can probe FFPs with smaller dwarf sources thanks to their superior photometric performance (high precision and stability) and diffraction-limited resolutions. As described in \S~\ref{sec:Obs}, we choose two representative types of sources (G2V and M0V).
\subsection{G-dwarf source}
\label{sec:gdwarf}
\begin{figure}[ht]
\begin{minipage}[b]{0.495\linewidth}
	\centering 
   \includegraphics[width=\linewidth]{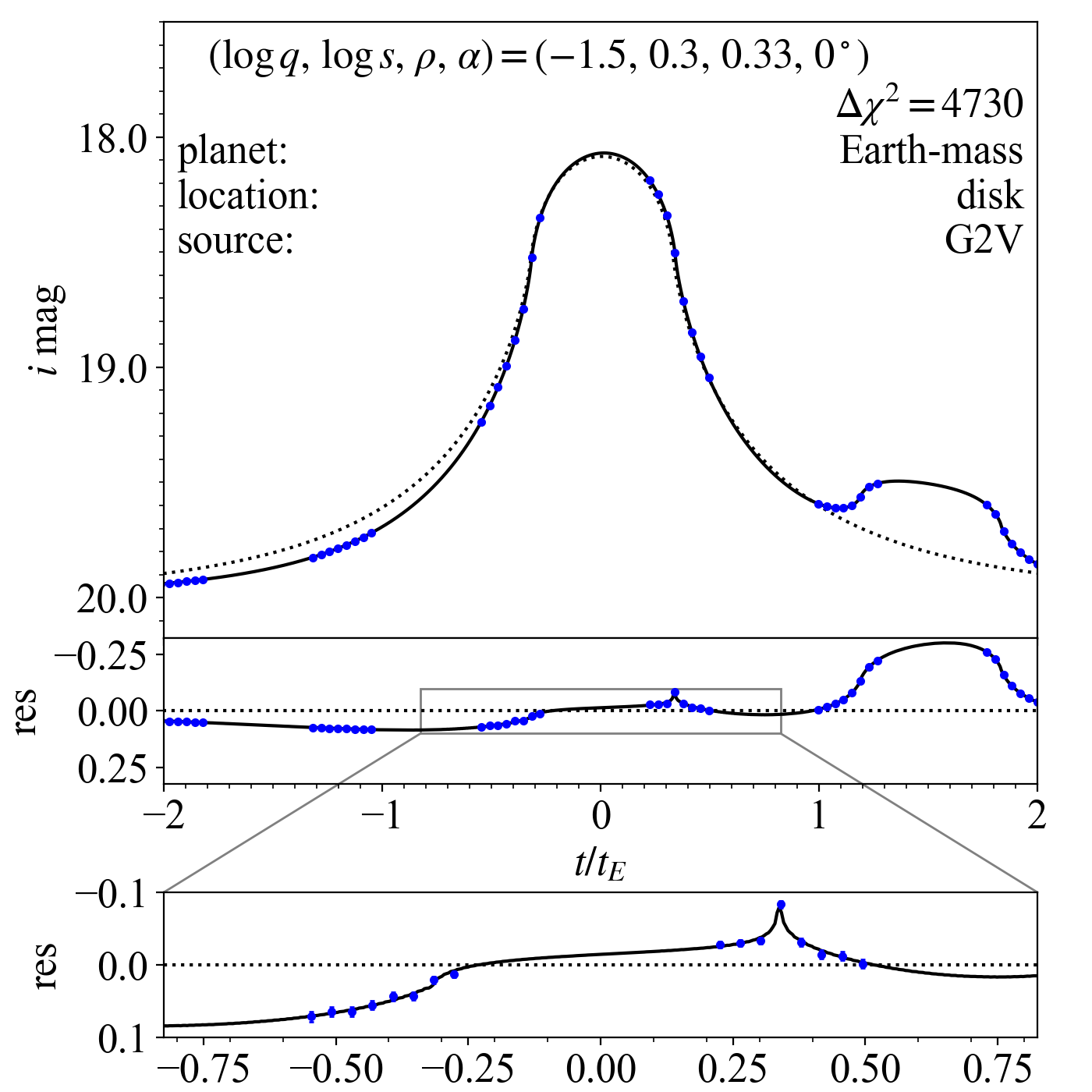}
    \end{minipage}%
\hfill    
    \begin{minipage}[b]{0.495\linewidth}
	\centering 
    \includegraphics[width=\linewidth]{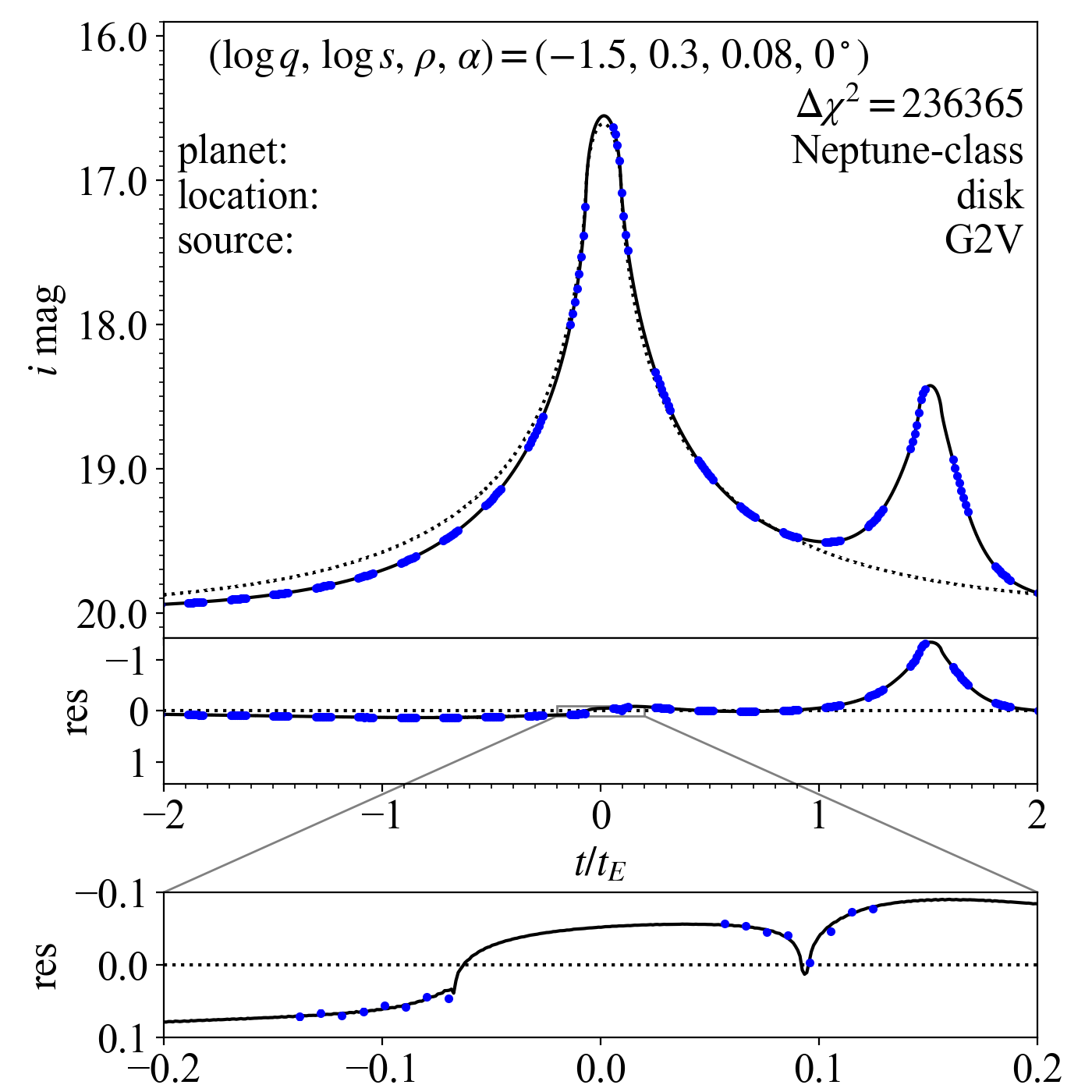}
   \end{minipage}%

\caption{Examples of simulated \csst\ 2L1S  light curves of a G2V source with planet-satellite parameters of $(\log{s},\log{q}, \alpha)=(0.3,-1.5, 0^{\circ})$. 
Left: The primary lens is an Earth-mass planet in the Galactic disk $(t_\E=0.090\,{\rm d}, \rho=0.33)$. The top sub-panel shows the simulated  data (blue dots), the underlying 2L1S model (black curve), and the best-fit 1L1S model (dotted curve). The middle sub-panel presents the residuals to the best-fit 1L1S model, and the zoomed-in view of the residuals around the light-curve peak is shown in the bottom sub-panel. Right: Similar to the left, but for a Neptune-class planet $(t_\E=0.357\,{\rm d}, \rho=0.08)$.}
	\label{fig:lc}
\end{figure}

\begin{figure}[h]
	\begin{minipage}[t]{0.499\linewidth}
	\centering 
    \includegraphics[width=\linewidth]{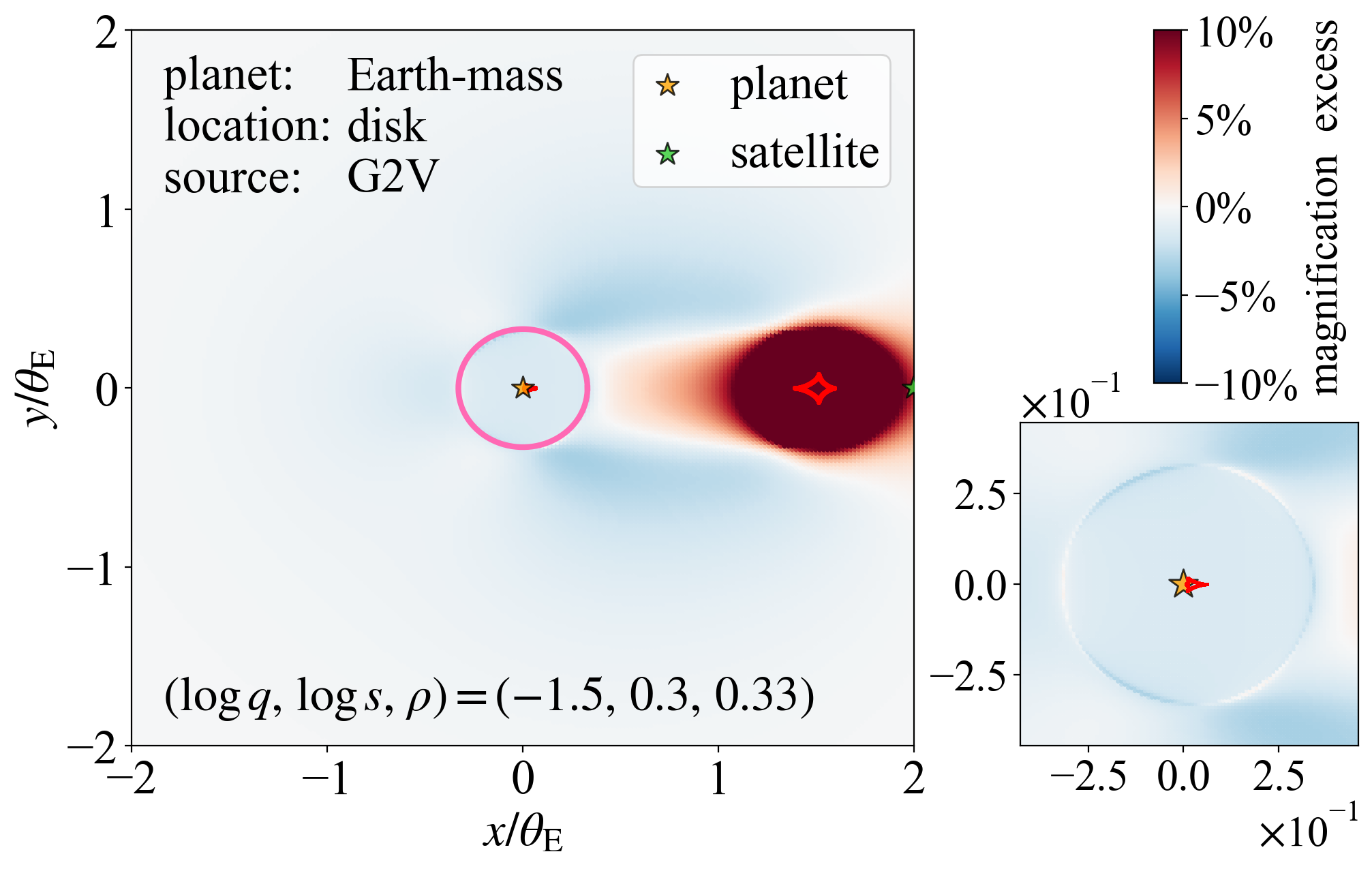}
    \end{minipage}%
    \hfill   
	\begin{minipage}[t]{0.499\linewidth}
	\centering 
    \includegraphics[width=\linewidth]{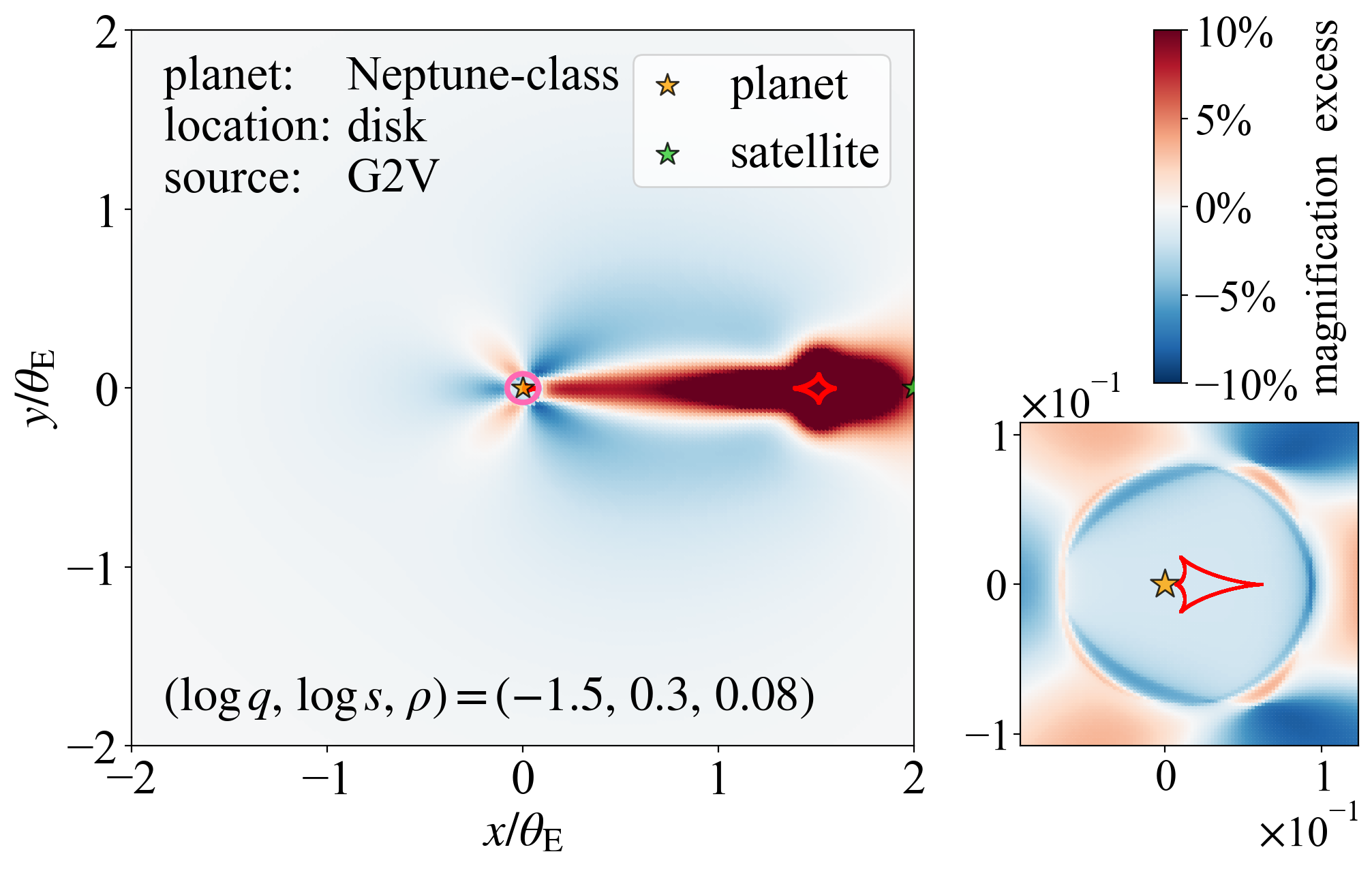}
    \end{minipage}%
    \hfill   
	\begin{minipage}[t]{0.499\linewidth}
	\centering 
    \includegraphics[width=\linewidth]{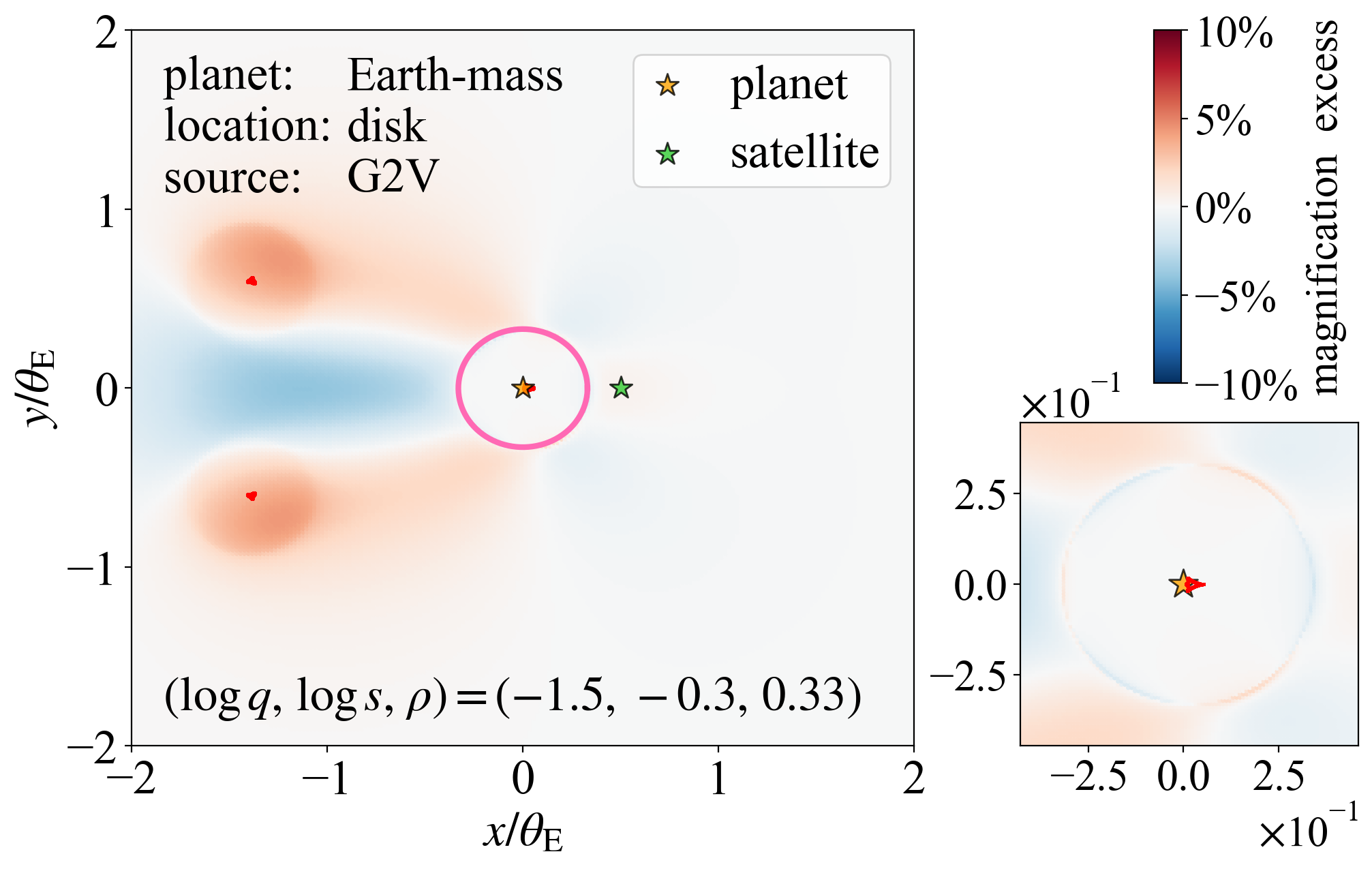}
    \end{minipage}%
    \hfill   
	\begin{minipage}[t]{0.499\linewidth}
	\centering 
    \includegraphics[width=\linewidth]{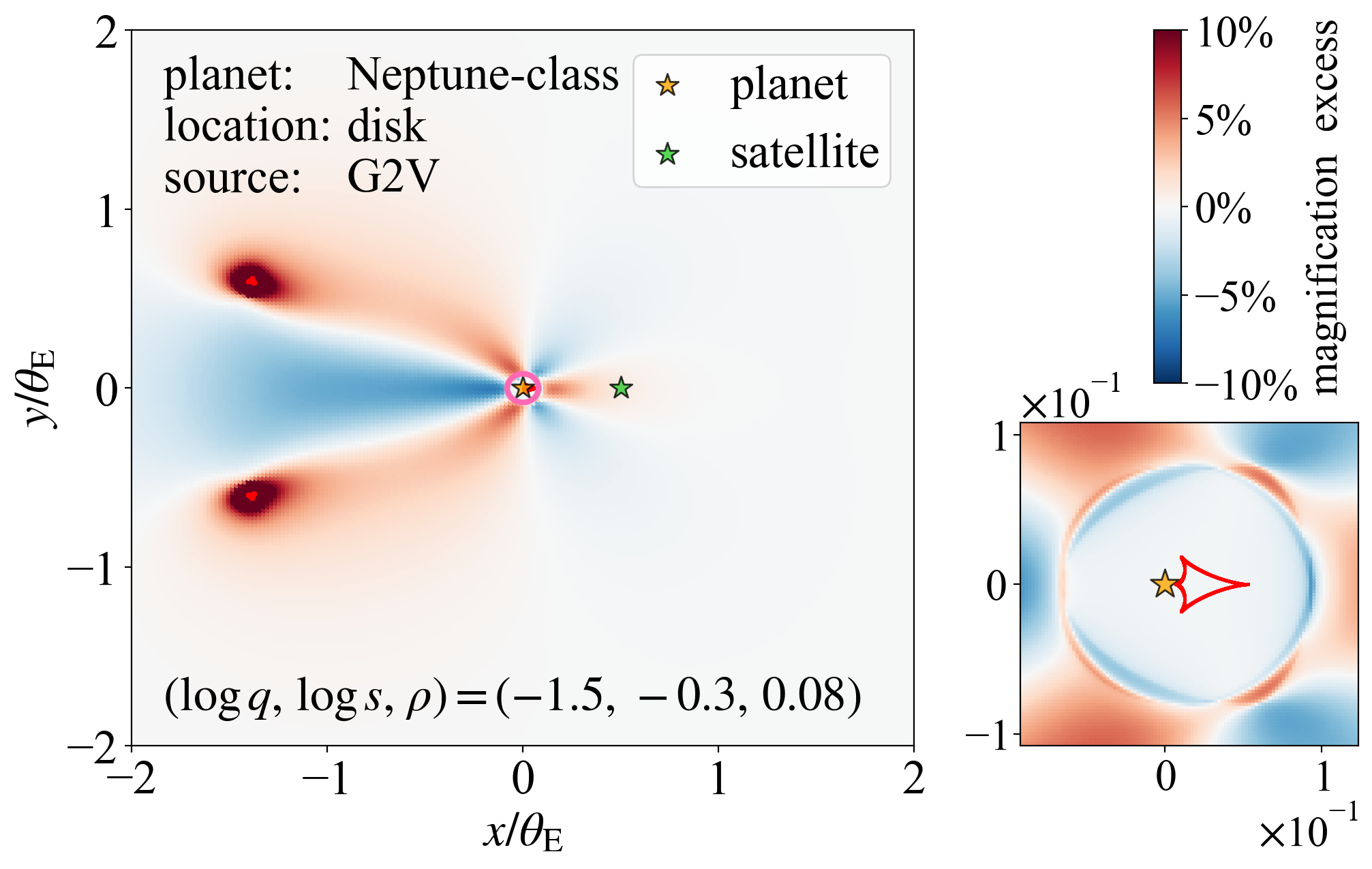}
    \end{minipage}%
	\caption{
The magnification excess maps for planet-satellite configurations with $\log{q}=-1.5$ and $|\log{s}|=0.3$. The color coding indicates fractional magnification differences between the 2L1S and 1L1S models. The maps with wide-separation ($\log{s}=0.3$) and close-separation ($\log{s}=-0.3$) cases are displayed in the upper and lower panels, respectively. The left and right panels correspond to Earth-mass and Neptune-class planets. The $x$-$y$ coordinates denote the source positions in units of $\theta_\E$, with the positive $x$-axis aligned in the direction pointing from the planet (gold star) to the satellite (cyan star). Magenta circles indicate the source stars' sizes ($\rho=0.33$ for Earth-mass lenses and $\rho=0.08$ for Neptune-class). Caustics are depicted in red, with the central caustic located near the planet and the satellite caustics positioned farther away. Each panel includes a lower-right inset offering a zoom-in view of the central caustics.}
	\label{fig:fig:excess_G2V}
\end{figure}
\begin{figure}[ht]
\begin{minipage}[b]{0.495\linewidth}
	\centering 
   \includegraphics[width=\linewidth]{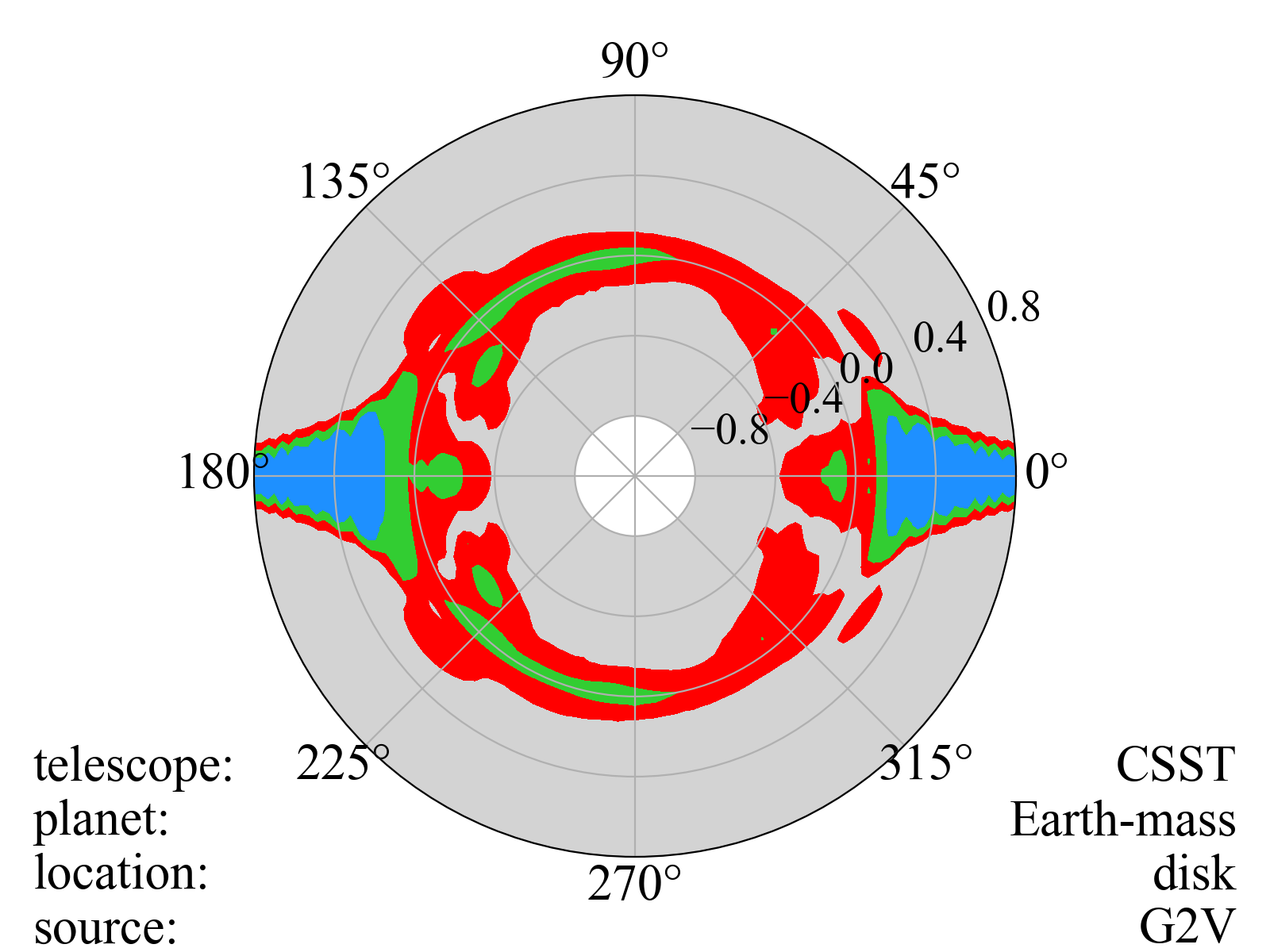}
    \end{minipage}%
\hfill    
    \begin{minipage}[b]{0.495\linewidth}
	\centering 
    \includegraphics[width=\linewidth]{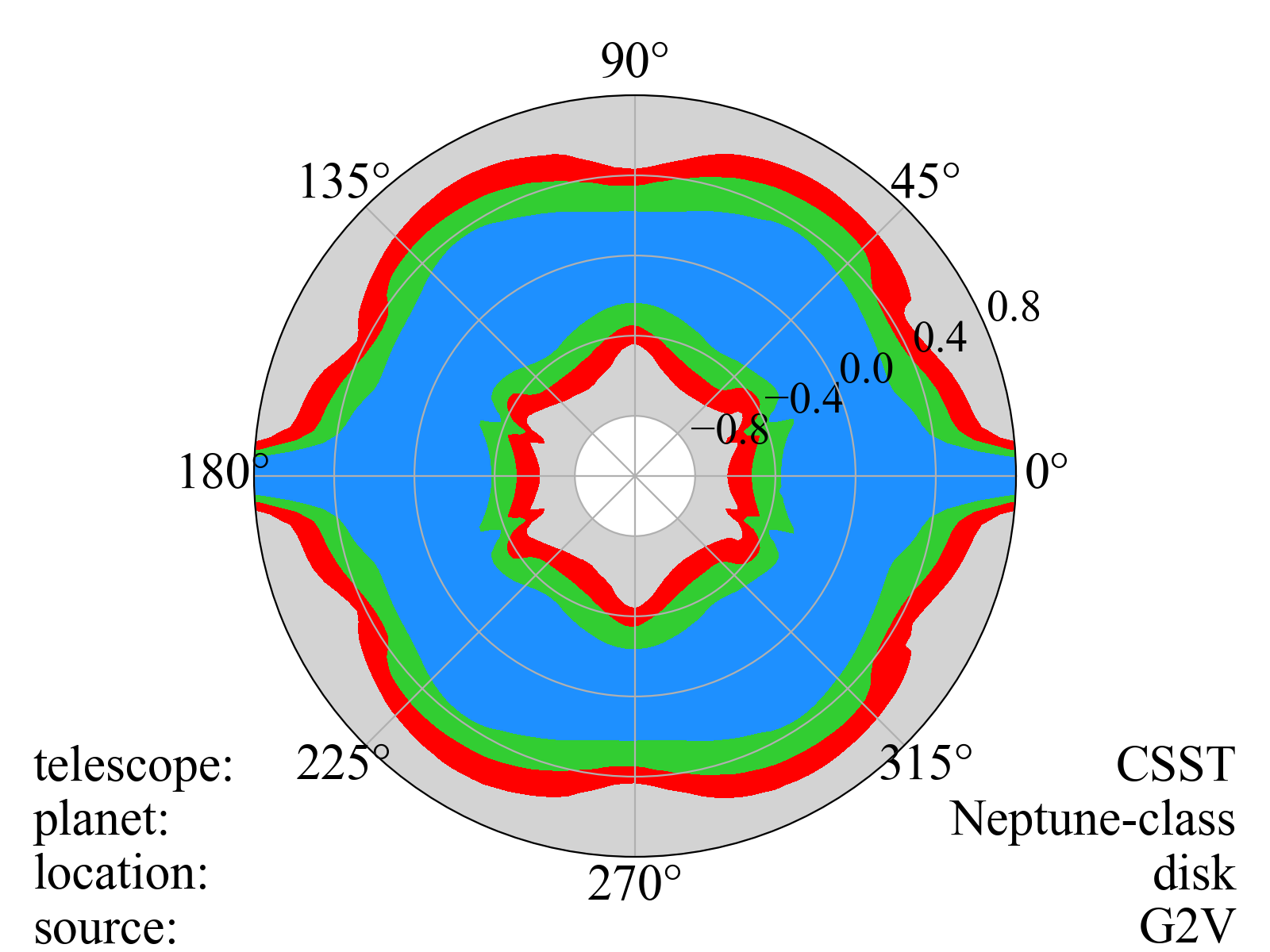}
   \end{minipage}%
	\caption{
	 $\Delta\chi^2$ distributions in polar coordinate $(\log{s},\alpha)$ at $\log{q}=-1.5$ (left: Earth; right: Neptune). Regions satisfying $\Delta\chi^2>100$, $>400$, and $>900$ are color-coded in red, green, and blue, respectively, and the satellites are detected in these regions according to our detection threshold ($\Delta\chi^2>100$). The satellites are undetected in the grey regions.
	}
	\label{fig:deltachi2}
\end{figure}

\begin{figure}[h]
	\begin{minipage}[t]{0.499\linewidth}
	\centering 
    \includegraphics[width=\linewidth]{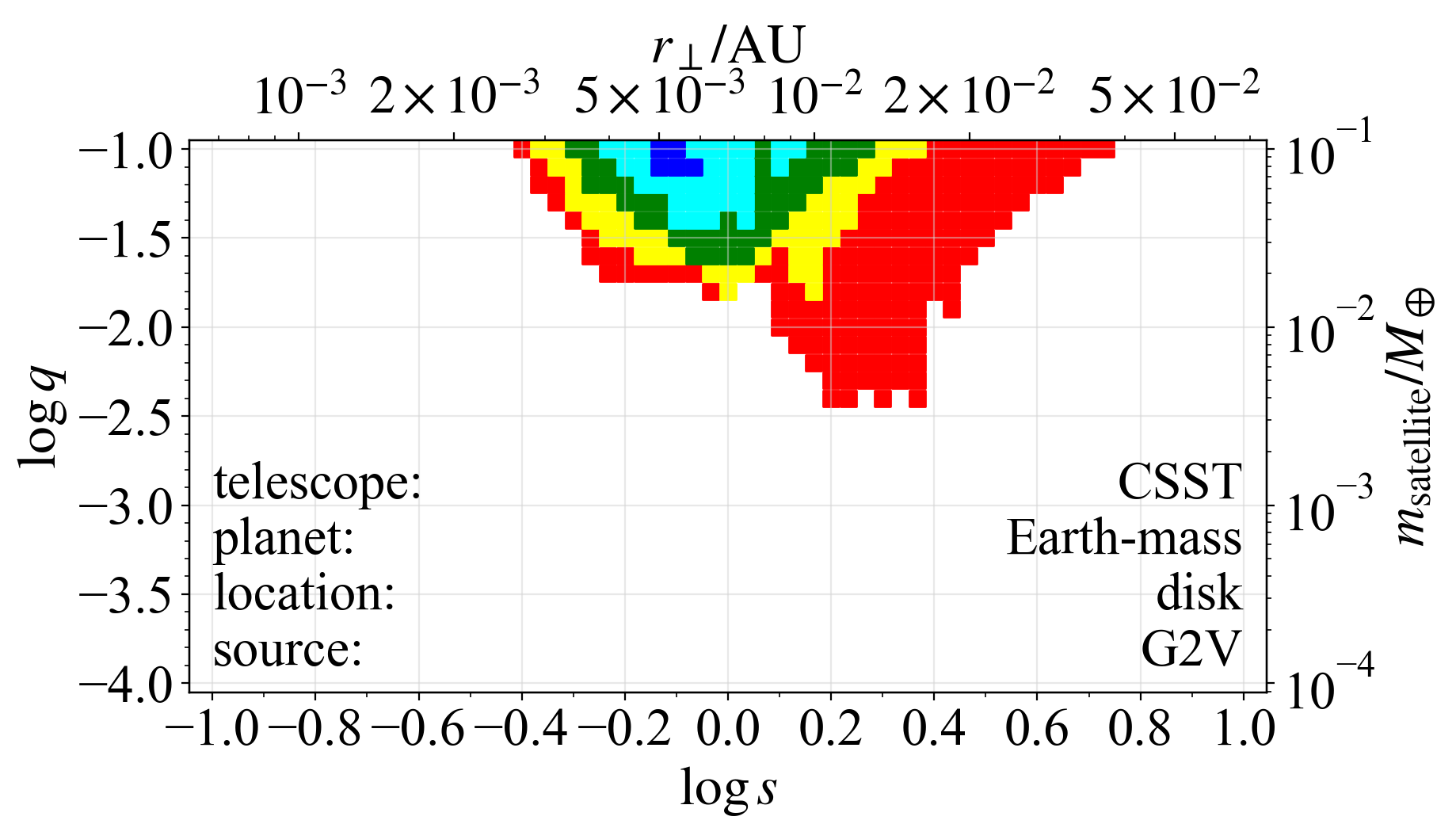}
    \end{minipage}%
    \begin{minipage}[t]{0.499\linewidth}
	\centering 
    \includegraphics[width=\linewidth]{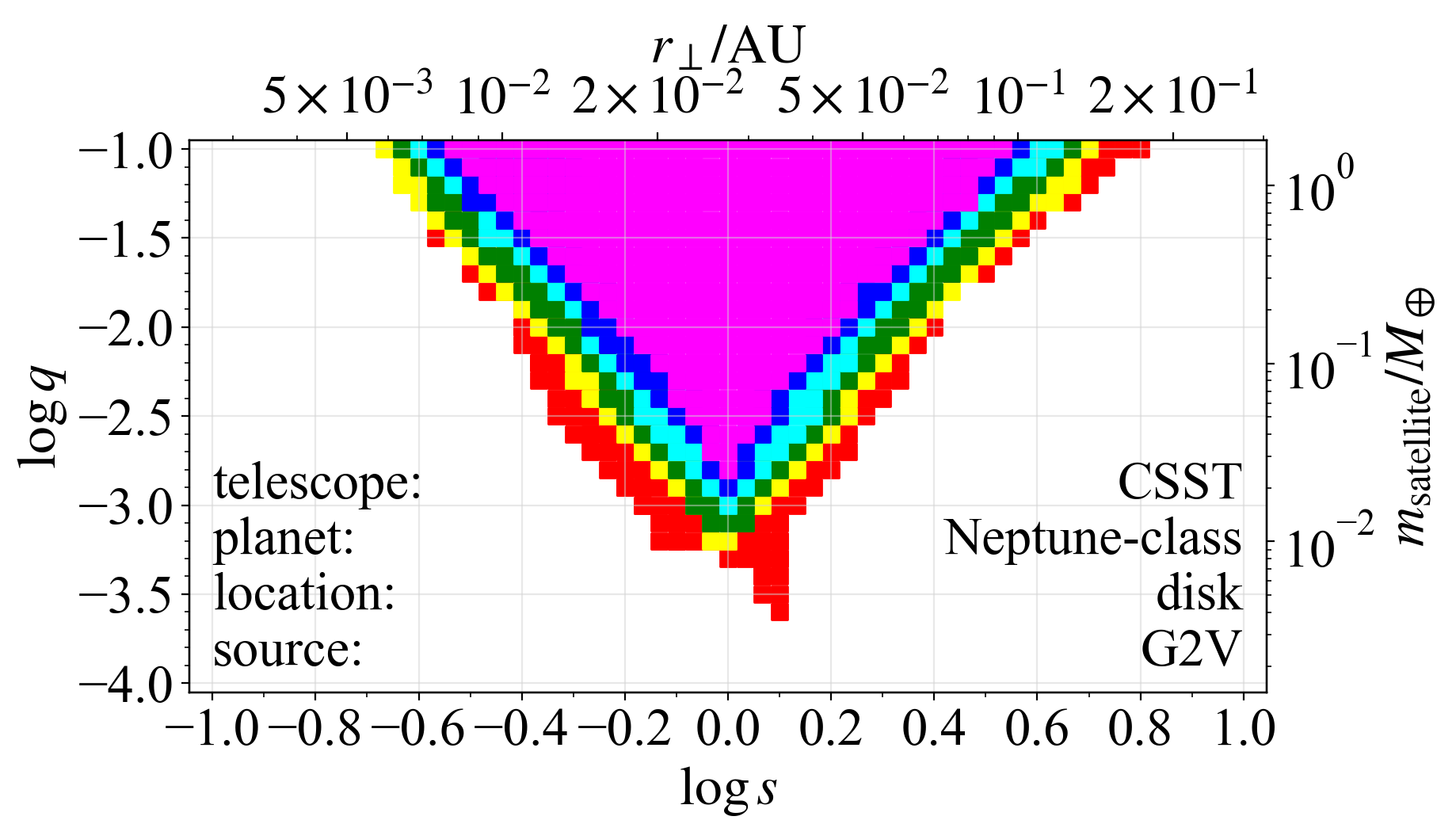}
    \end{minipage}%
    
    \begin{minipage}[t]{0.499\linewidth}
	\centering 
    \includegraphics[width=\linewidth]{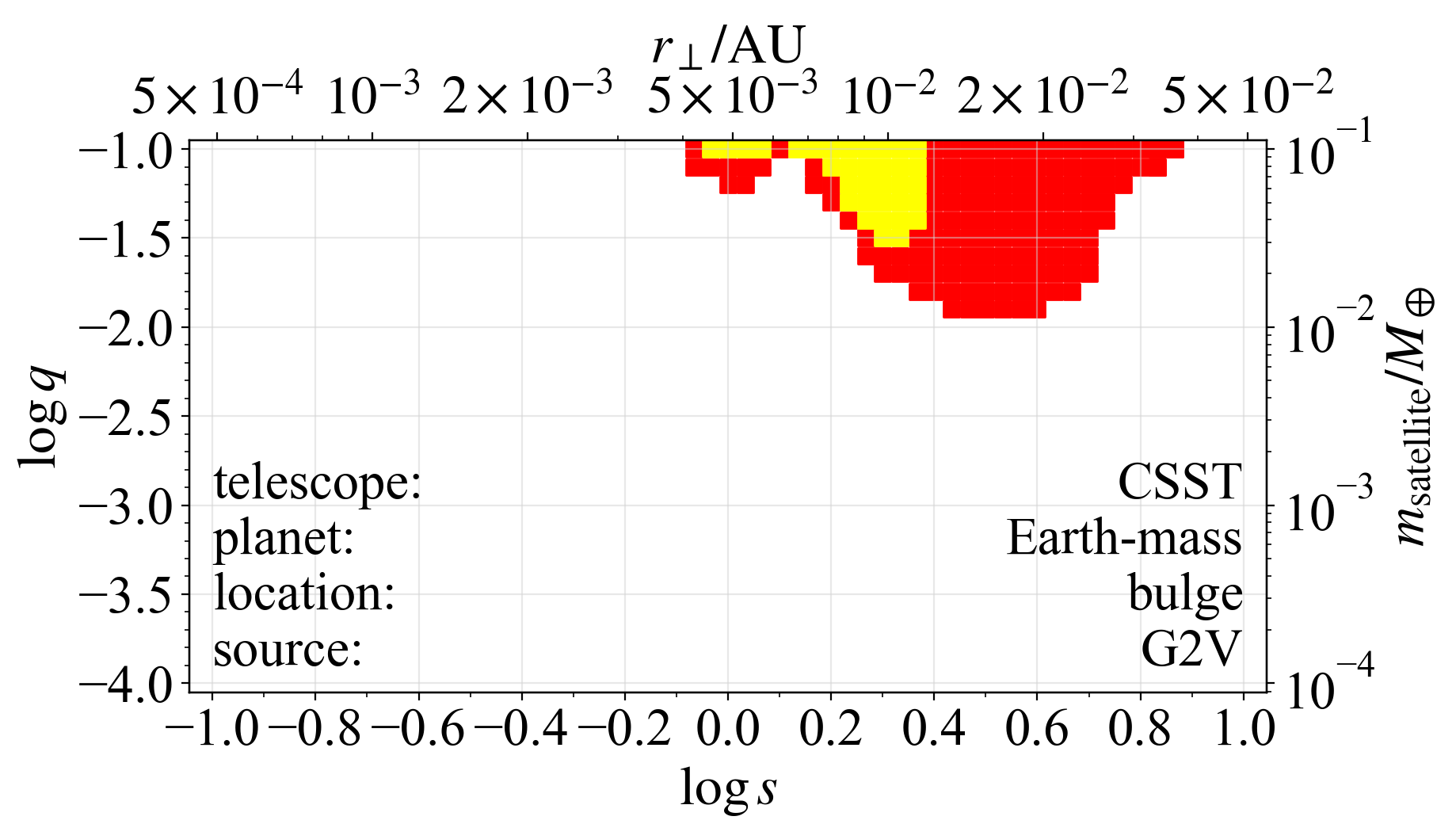}
    \end{minipage}%
    \begin{minipage}[t]{0.499\linewidth}
	\centering 
    \includegraphics[width=\linewidth]{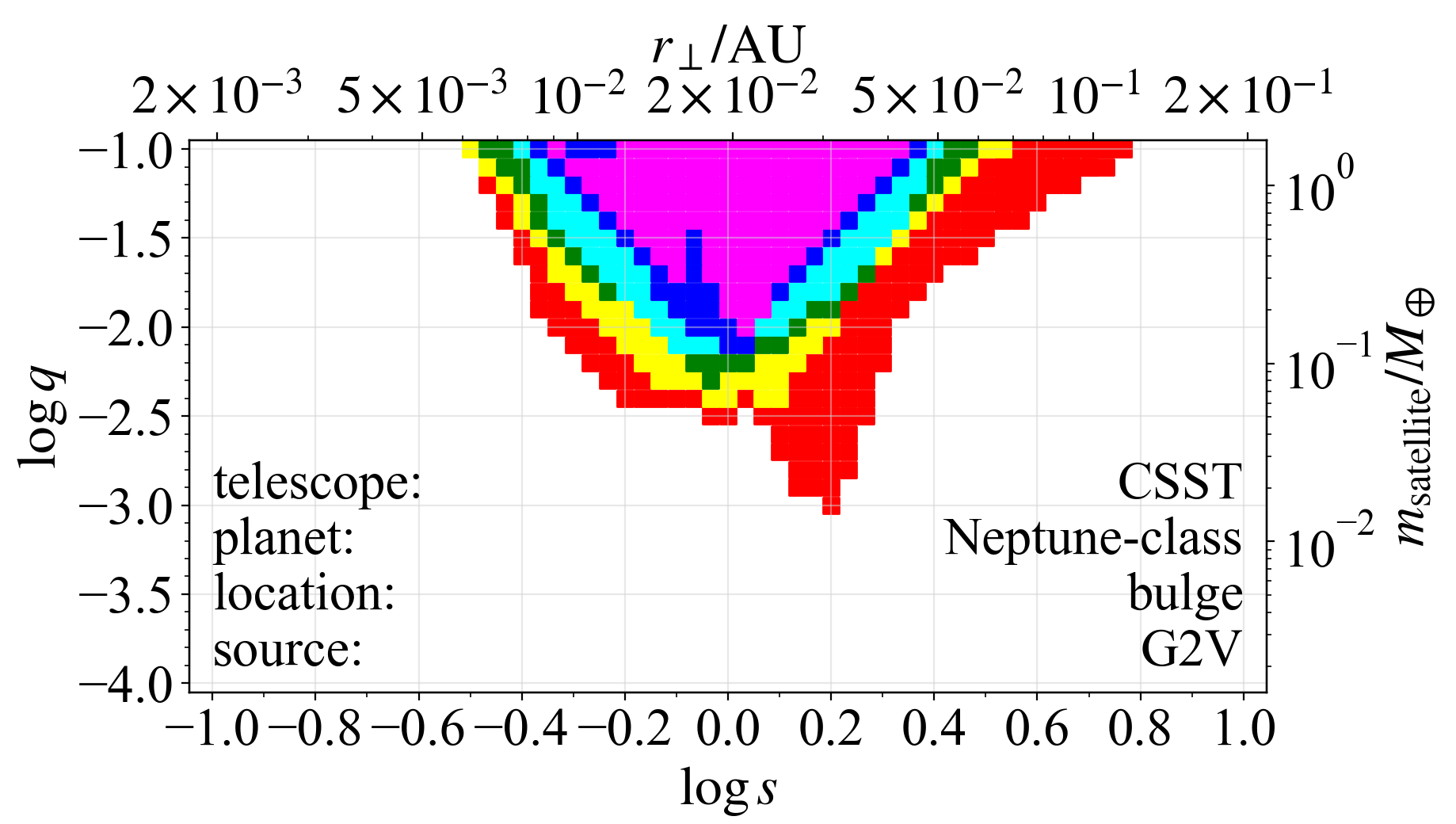}
    \end{minipage}%
	\caption{\csst\ satellite detection efficiency as a function of the satellite-planet mass ratio $q$ and planet-satellite projected separation $s$, with a G2V source. The satellite mass $m_{\rm satellite}$ and physical projected separation $r_\perp$ are also displayed. The Earth-mass (left panels) and Neptune-class (right panels) planetary lenses are located in the disk (upper panels) and bulge (lower panels), respectively. The color coding indicates efficiency levels: 10\% (red), 25\% (yellow), 50\% (green), 75\% (cyan), 90\% (blue) and 100\% (magenta). }
	\label{fig:G2V_eff}
\end{figure}

We first discuss results involving a G2V source. Figure~\ref{fig:lc} shows 
two examples of mock \csst\ light curves for detectable satellites with $\approx3\%$ of planetary mass ($\log q = -1.5$). In both cases, the planets are in the Galactic disk, with Earth-mass (left) and Neptune-class (right), respectively. One class of satellite signals is the ``central-caustics''  \citep{Griest98} perturbations near the peak (see the bottom-left sub-panels). Central-caustic perturbations exist for essentially all position angles of the source trajectories. There are also off-peak perturbations due to caustics along the binary axis but away from the position of the primary. These caustics are called ``planetary caustics'' in the context of the star-planet lens system, and here we refer them as ``satellite caustics'' for clarity. Depending on whether $s>1$ (wide) or $s<1$ (close) , there are one or two satellite caustics, respectively on the same and or opposite side of the satellite with respect to the planet. These two examples are both wide with  $s=2.0\,(\log s = 0.3)$. In contrast to central caustics, considerable perturbations are usually produced when the source trajectories cross or approach the caustic. For the examples, we pick source trajectories with $u_0=0$ and $\alpha=0^\circ$, which cross the satellite caustics.

We further explore the satellite perturbations using the magnification excess maps \citep[see, e.g.,][]{Dong09}, which display the distributions of fractional differences in magnifications between the 2L1S models (shifted to the center of the magnification pattern, \citealt{Yoo04b, Dong06}) and the underlying 1L1S models. See Figure~\ref{fig:fig:excess_G2V} for the magnification excess maps of Earth-mass and Neptune-class lenses with $\log{q}=-1.5$ and $|\log{s}|=0.3$ (i.e., both wide and close). The satellite signals are concentrated in the vicinities of the caustics (red). The satellite caustics show substantially more prominent perturbations than the central caustics (highlighted in the lower-right insets of the sub-panels), and the most significant perturbations are from the satellite caustics of the wide-separation cases (upper). The signals are generally stronger for the Neptune-class planets (right) than the Earth-mass planets (left). The signals for the Earth-mass planet are more ``washed out'' due to the larger source size $(\rho=0.33)$ than that for the Neptune $(\rho=0.08)$, while the larger finite-source effects also considerably broaden the perturbation region surrounding the satellite caustics of the wide case for the Earth-mass planet.

Figure~\ref{fig:deltachi2} presents the  $\Delta\chi^2$ distributions as functions of $\log s$ and $\alpha$ for mock  data with $\log{q}=-1.5$. The regions with detected satellites are significantly larger for Neptune-class FFPs (left) compared to Earth-mass FFPs (right), primarily due to the smaller source sizes for the simulated Neptune-class FFP cases, as discussed above.

The top-right panel of Figure~\ref{fig:G2V_eff} displays the detection efficiency of satellites orbiting a Neptune FFP in the disk, exhibiting a triangular pattern that is symmetric with respect to $s=1$ $(\log s=0)$. Such a pattern is characteristic of planetary detection efficiency in high-magnification events \citep[see, e.g.,][]{Dong06, Gould10}, arising from the close-wide ($s$-$s^{-1}$) degeneracy of the central-caustic perturbations \citep{Griest98, Dominik99, Bozza99, An05, An21}. The sensitivity zone reaches down to $q\approx10^{-3}$ near the Einstein radius, corresponding to a Moon-mass ($\sim 10^{-2}M_\oplus$) satellite at physical projected separation $r_\perp=sR_\E\approx3\times10^{-3}$\,\au. For Earth-mass satellites, the sensitivity covers $\sim1$\,dex around the Einstein radius, corresponding to  $r_\perp \approx 0.01\text{--}0.1$\,\au. 

In comparison, the detection efficiency of satellites for an Earth-mass FFP (top-left panel of Figure~\ref{fig:G2V_eff}) shows a more asymmetric pattern, with substantially extended sensitivity at $s>1$ due to satellite-caustic perturbations broadened by the finite-source effects of the relatively large $\rho$. There is modest sensitivity ($\sim 10\%$ efficiency) to Moon-mass satellites ($q\approx0.01$) at $s\approx1.5\text{--}2.5$ ($r_\perp \sim 0.01$\,\au). For satellites with about ten times Moon mass ($q\approx0.1$), the $50\%$ sensitivity zone encompasses $0.5\lesssim s \lesssim 2$, while the $10\%$ zone extends from $s\sim0.4$ to $s\sim7$. 

The bottom panels of Figure~\ref{fig:G2V_eff} display the results for FFPs in the bulge, showing considerably smaller sensitivity zones than the disk FFPs. This difference arises from two main factors: bulge-lens events have larger $\rho$, which tends to reduce the amplitudes of the signals, and events with disk lenses have longer timescales, enabling better light-curve coverage.

\subsection{M-dwarf source}
\label{sec:mdwarf}

\begin{figure*}[h]
	\begin{minipage}[t]{0.499\linewidth}
	\centering 
    \includegraphics[width=\linewidth]{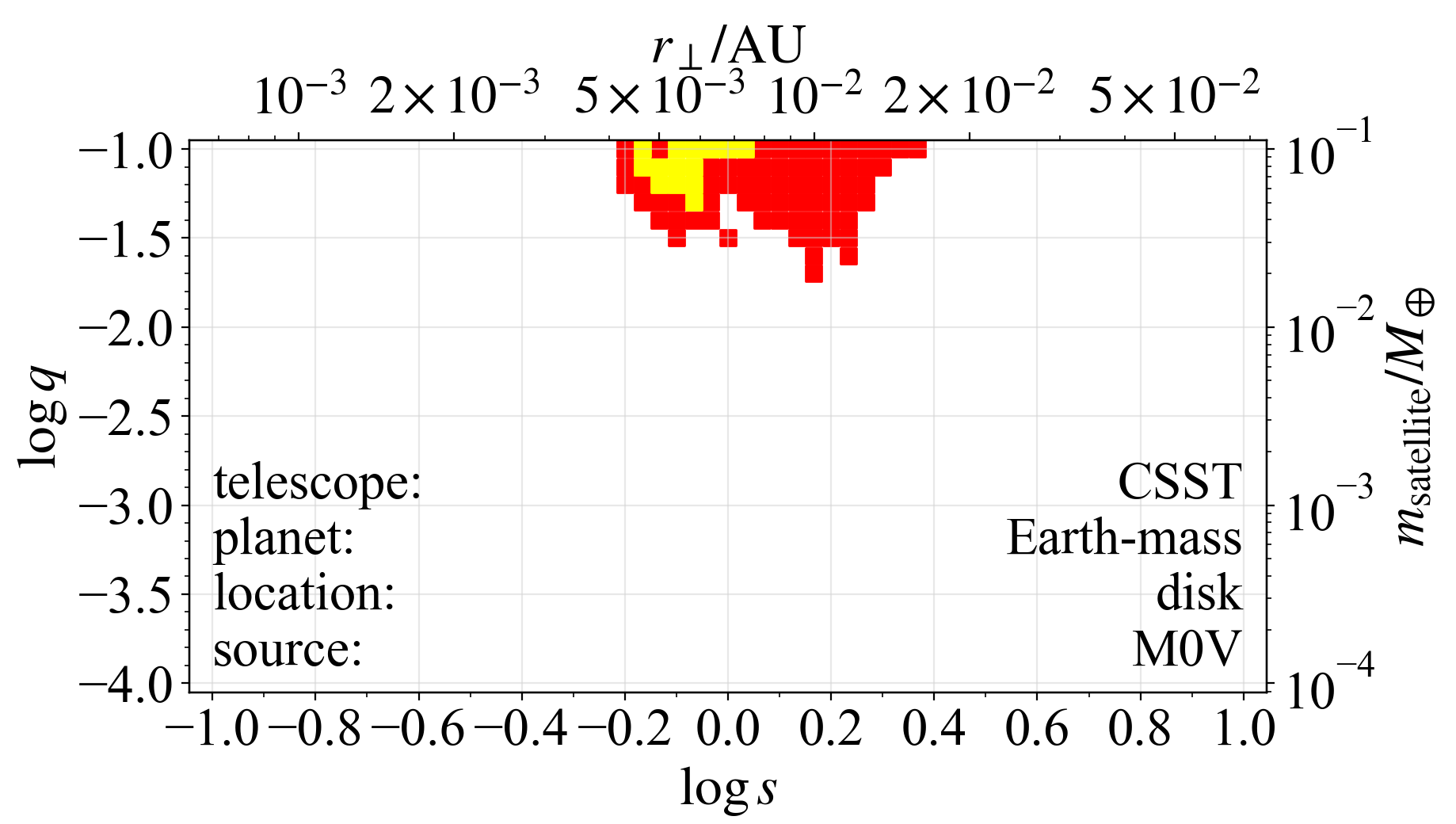}
    \end{minipage}%
    \begin{minipage}[t]{0.499\linewidth}
	\centering 
    \includegraphics[width=\linewidth]{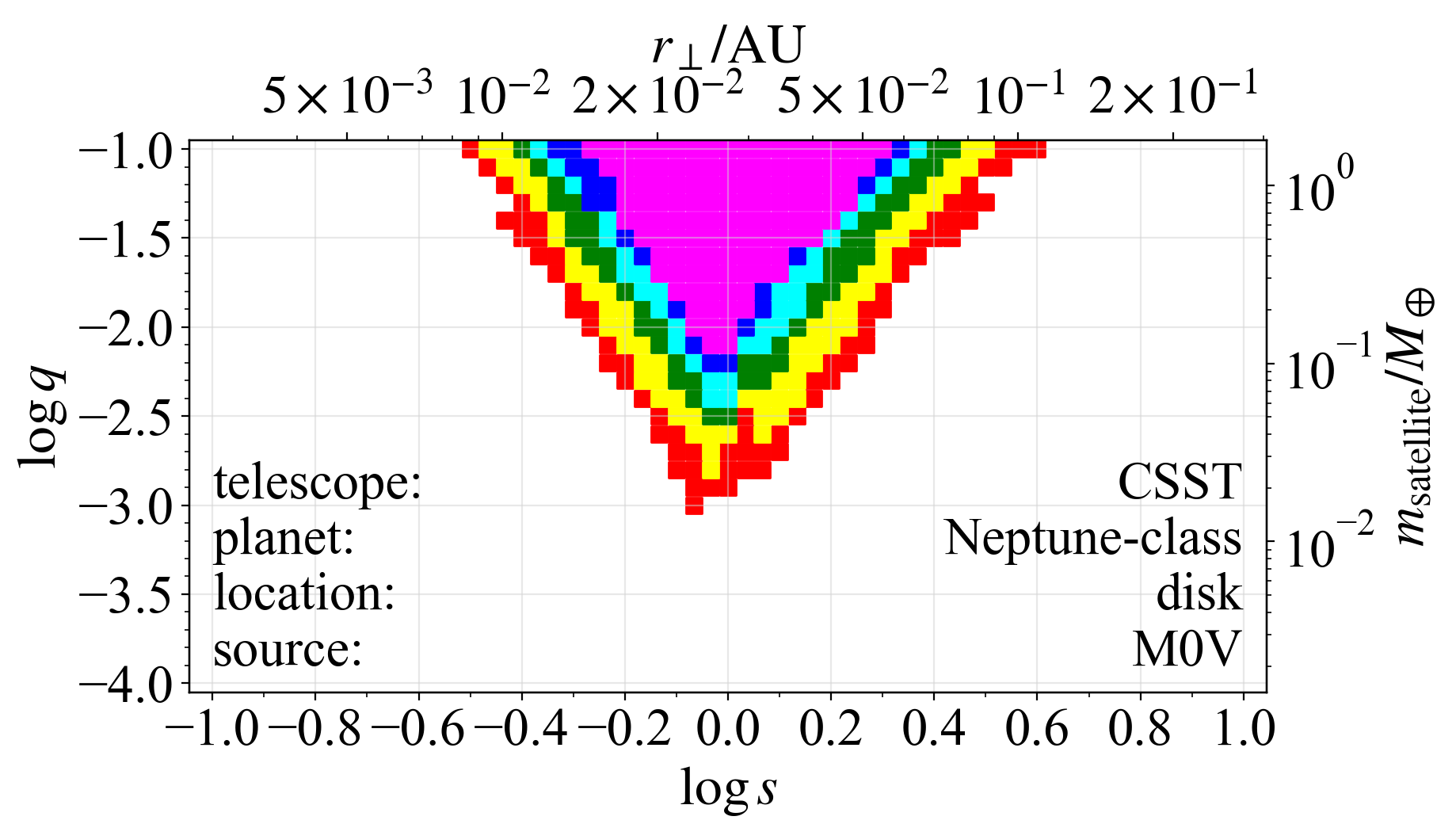}
    \end{minipage}%
	\caption{Satellite Detection efficiency for FFPs (left: Earth; right: Neptune) in the disk with an M0V source.}
	\label{fig:Mdwarf_eff}
\end{figure*}

We evaluate \csst\ satellite detection efficiency using an M0V star as the source star, shown in Figure~\ref{fig:Mdwarf_eff}. In comparison to a G2V star, an M0V source is significantly fainter, making the satellite signals more difficult to detect. Additionally, the smaller size of the source further reduces the sensitivity zone at $s>1$. 
The left panel of Figure~\ref{fig:Mdwarf_eff} illustrates the results for an Earth-mass FFP lens in the disk with an M0V source, revealing limited sensitivity, capable of detecting only satellites with relatively large mass ratios ($q \sim 0.1$) near $s \sim 1$.
For a Neptune-class lens (right panel of Figure~\ref{fig:Mdwarf_eff}), the sensitivity zone in $\log q$ is still substantial, but it is no longer sensitive to Moon-mass satellite ($\sim 0.01\,M_\oplus$).

\subsection{\rst}
\label{sec:roman}
\begin{figure*}[h]
	\begin{minipage}[t]{0.499\linewidth}
	\centering 
    \includegraphics[width=\linewidth]{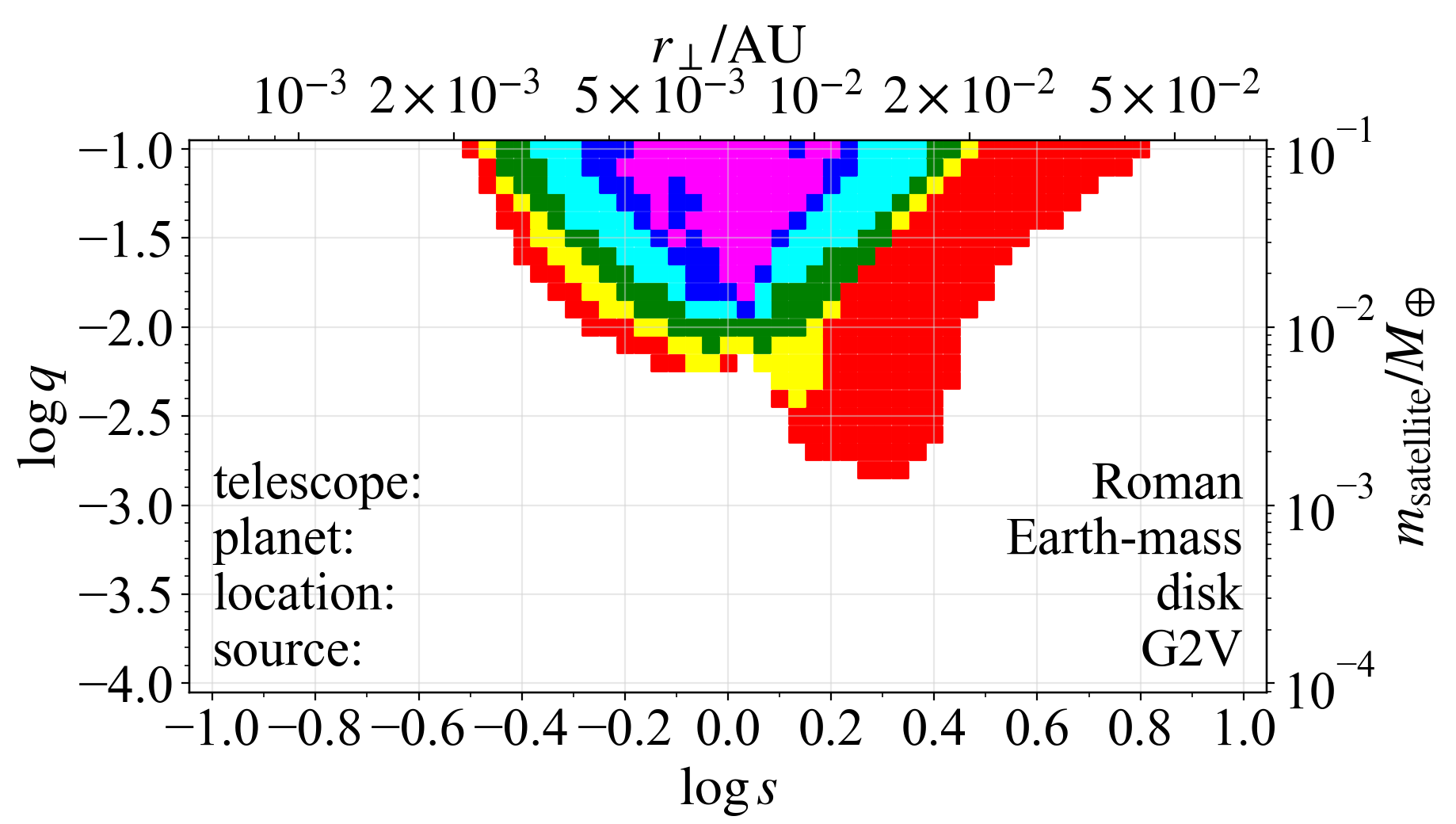}
    \end{minipage}%
        \begin{minipage}[t]{0.499\linewidth}
	\centering 
    \includegraphics[width=\linewidth]{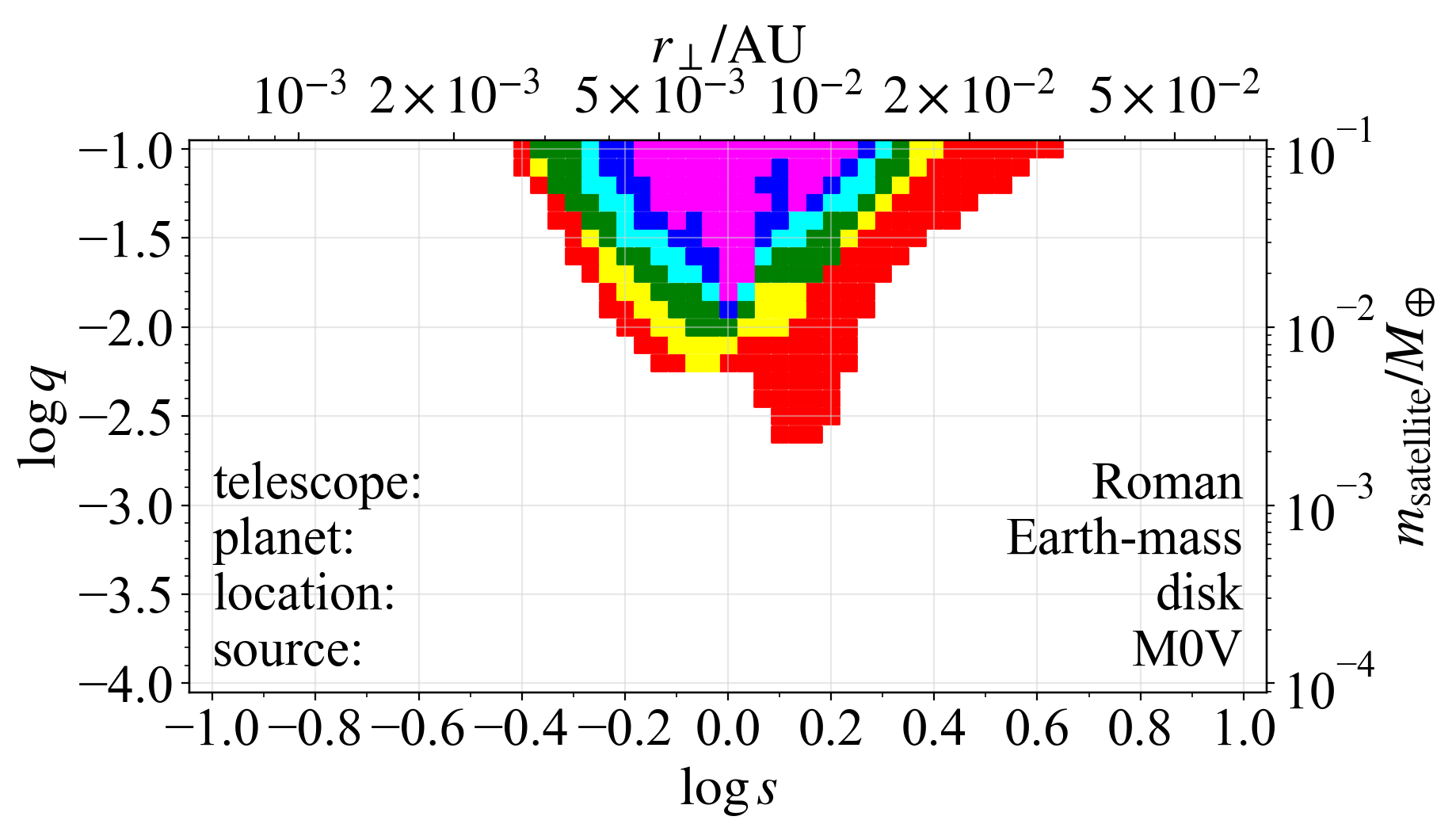}
    \end{minipage}%
	\caption{\rst\, satellite sensitivity for Earth-mass FFP events with G2V (left) and M0V (right) sources.}
	\label{fig:Roman_eff}
\end{figure*}

\begin{figure*}[h]
	\begin{minipage}[t]{0.499\linewidth}
	\centering 
    \includegraphics[width=\linewidth]{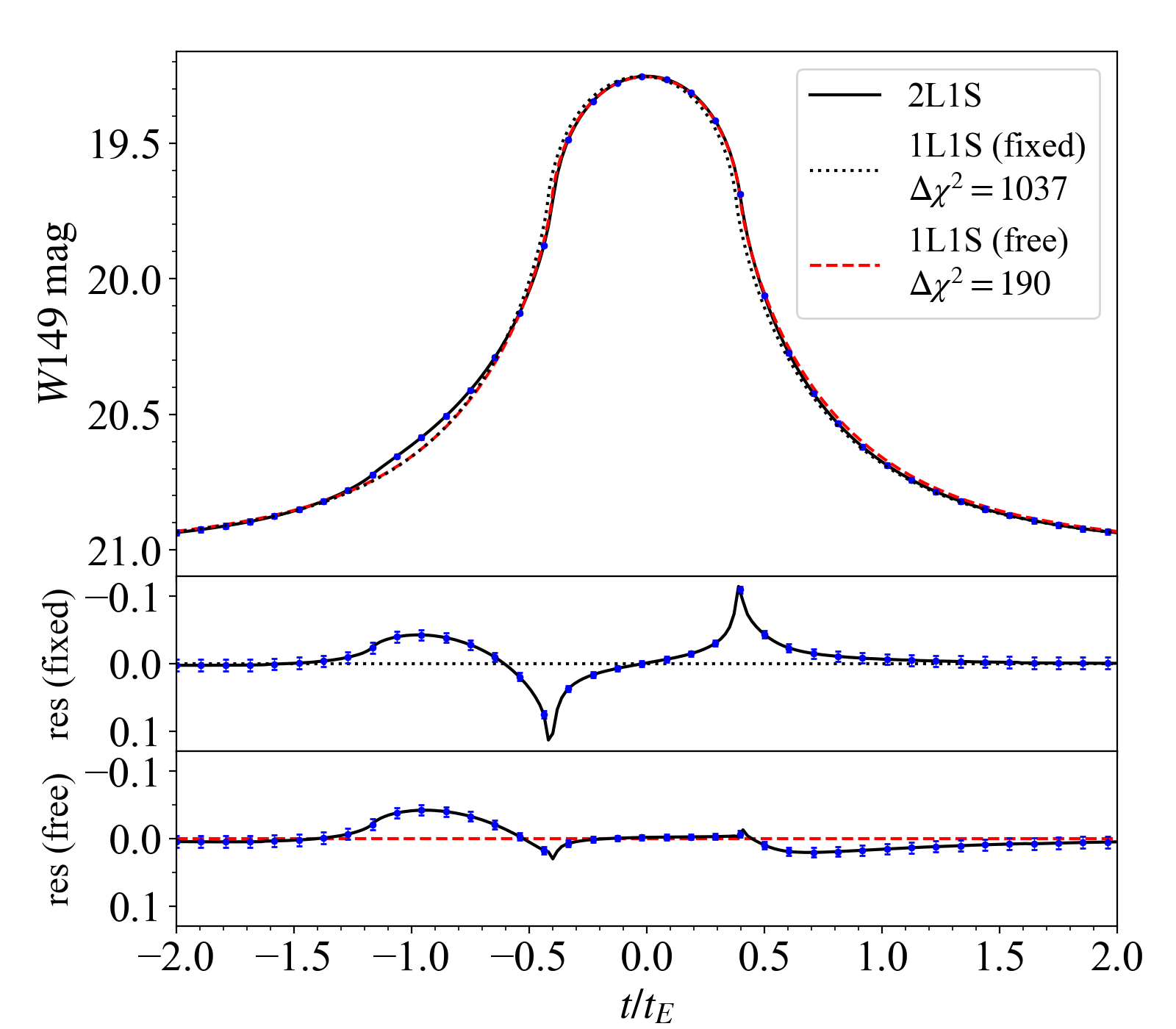}
    \end{minipage}%
    \begin{minipage}[t]{0.499\linewidth}
	\centering 
    \includegraphics[width=\linewidth]{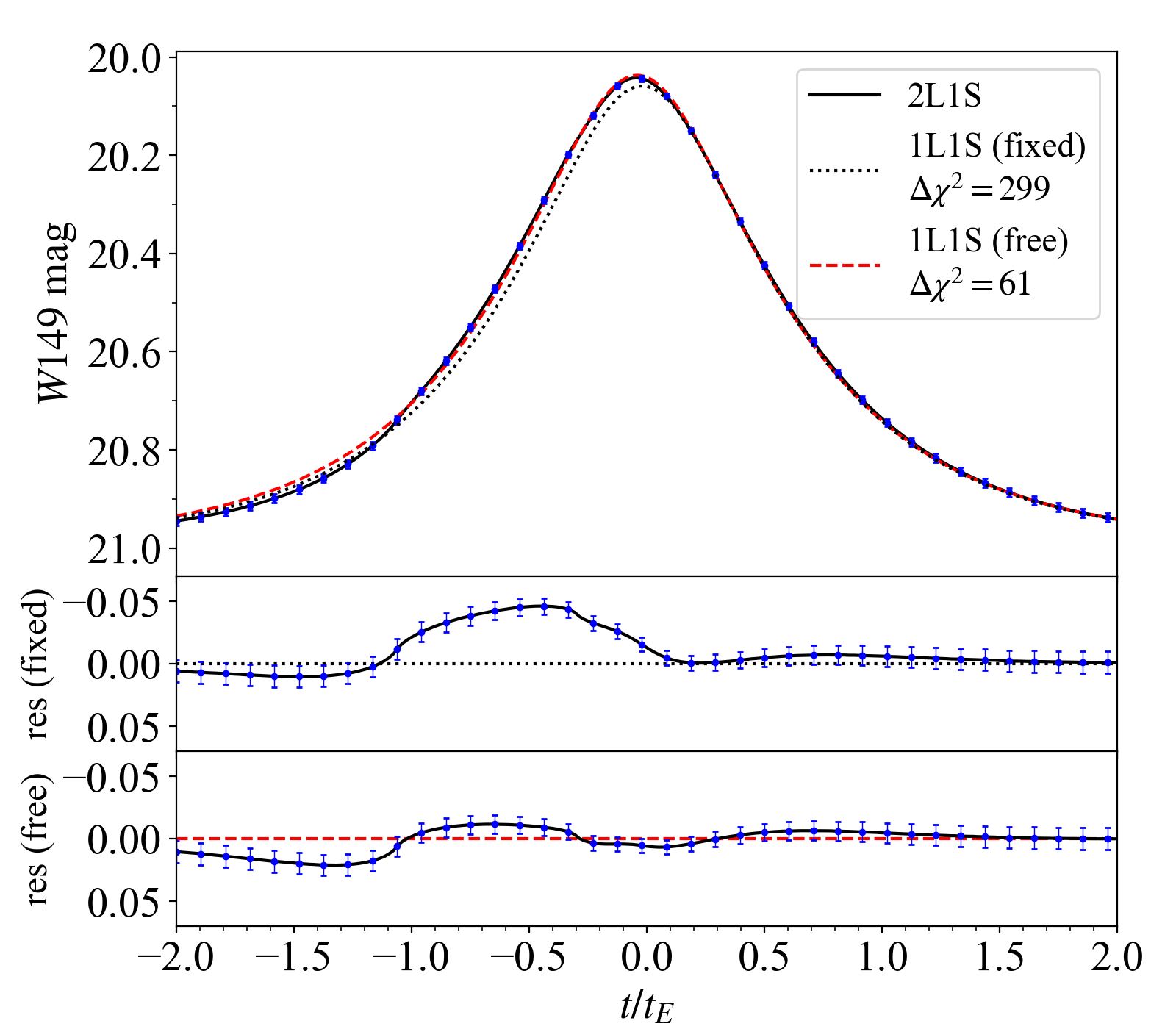}
    \end{minipage}%
    
	\caption{Two examples as shown in the (a) and (b) panels of Figure~1 in \cite{Sajadian23}. Both share the same $t_\E=0.1$\,d, $q=0.03$, $s=0.7$, and $\rho=0.4$, while they differ in $u_0$ and $\alpha$. We generate the mock data (blue dots) based on the 2L1S models 
(solid black lines) and fit them to two sets of 1L1S models: fixed (black dotted lines) and free (red dashed lines) parameters. In the fixed-parameter set, the parameters are fixed to be identical to those underlying the 2L1S model, in the same fashion as in \cite{Sajadian23}. The free-parameter set allows the parameters to vary freely, yielding the best-fit $\Delta\chi^2$ values (as indicated in the legends) that are significantly smaller than the former set. The residuals to the fixed-parameter and best-fit free-parameter models are shown in the bottom two sub-panels, respectively. Note that \cite{Sajadian23} introduced randomized scatters to the data points, preventing exact reproduction of their $\Delta\chi^2$ values. However, this does not impact our primary conclusion regarding the comparison between fixed and free models.}
	\label{fig:compare} 
\end{figure*}

We compute the satellite detection efficiency for {\it Roman}. As shown in 
Figure~\ref{fig:Roman_eff}, \rst\, has greater satellite sensitivity than \csst. For an Earth FFP in the disk with a G2V source (the left panel of Figure~\ref{fig:Roman_eff}), \rst's sensitivity extends to Moon-mass satellites near $s=1$ and even some sub-Moon-mass (down to $\sim0.2\,M_{\rm Moon}$) satellites at wide separations ($s\sim2$, approximately $0.02$\,\au). In comparison, with a M0V source, {\it Roman}'s detection efficiency (the right panel of Figure~\ref{fig:Roman_eff}) decreases only modestly, unlike \csst's more drastic drop (the top-left panel of Figure~\ref{fig:G2V_eff} in compared to the left panel of Figure~\ref{fig:Mdwarf_eff}), as discussed in \S~\ref{sec:mdwarf}.
This is primarily because \rst\, observes in the infrared, which is significantly more sensitive to the M-dwarf spectral energy distribution than the optical \csst\  observations.

\citet{Sajadian23} investigated the detectability of FFPs' satellites with \rst, focusing on analogs to 25 planet-satellite pairs in the Solar System and evaluating systems with $\log q$ uniformly distributed over the range of $[-9,\,-2]$.  Their reported detection efficiency is markedly higher than that derived from our approach. \citet{Sajadian23} did not specify how they fitted the mock data with 1L1S models to calculate the $\Delta\chi^2$ values. We find that fixing the 1L1S parameters $(t_0, u_0, t_\E)$ to those of the underlying 2L1S 
models is needed to reproduce the difference between 2L1S and 1L1S models illustrated in the residual sub-panels of their Figure~2. As shown in Appendix B of \cite{Dong06}, this approach can significantly overestimate the detection efficiency of planets. Therefore, treating the 1L1S parameters as free parameters is necessary to properly simulate the detection process, as implemented in our analysis. Figure~\ref{fig:compare} provides two examples from \cite{Sajadian23}, both exhibiting significantly larger $\Delta\chi^2$ values than those using our procedure of freely fitting 1L1S models. Consequently, we infer that \cite{Sajadian23} overestimated their satellite sensitivity.

\begin{figure}[h]
	\centering
	\includegraphics[width=0.95\textwidth]{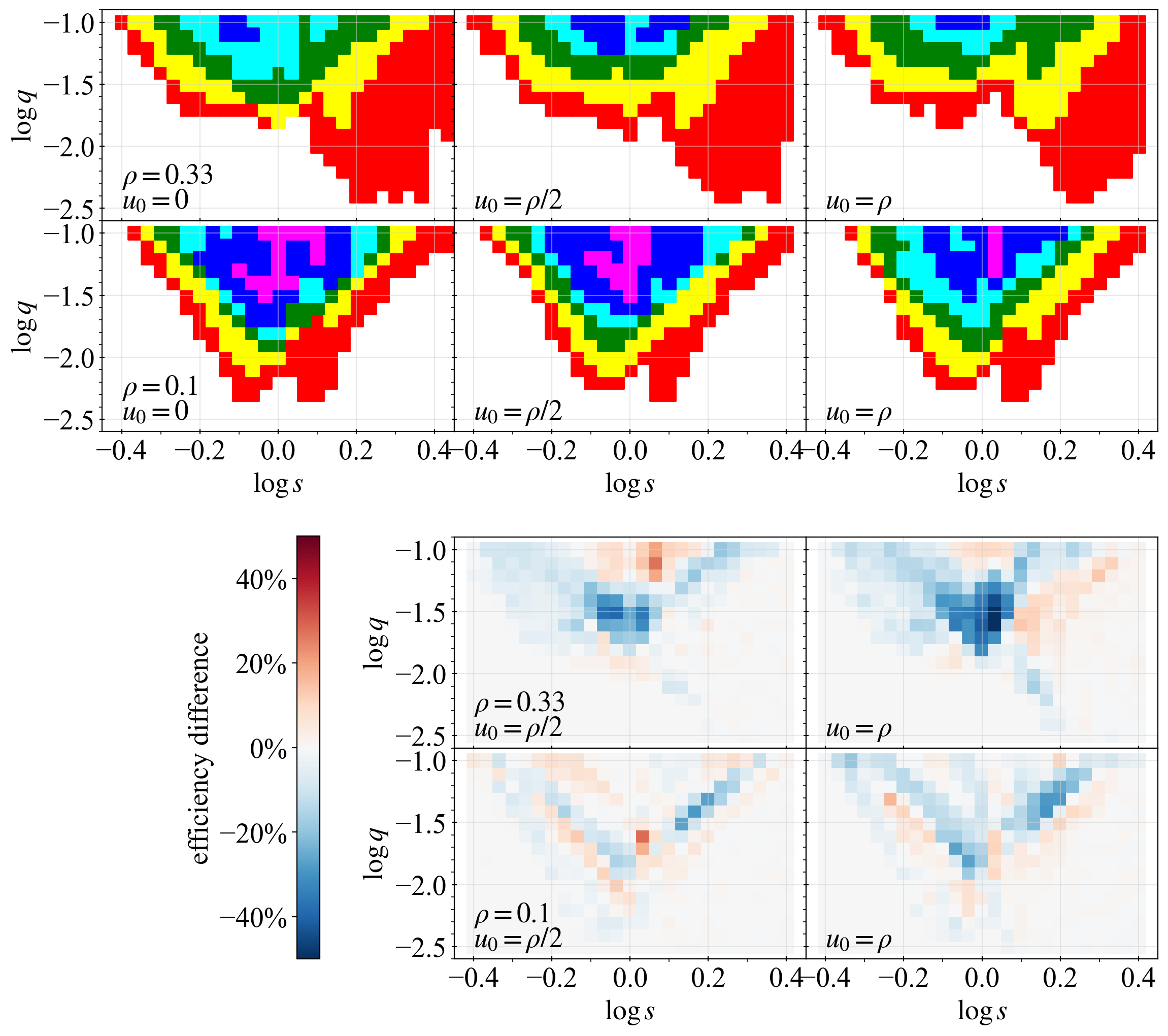}
	\caption{The first (second) row presents the \csst\ satellite detection efficiency in $s$-$q$ parameter space, with three different impact parameters $u_0=0,\,\rho/2,\,\rho$, for an Earth-mass (Neptune-class) lens in the disk with a G2V source (M0V source). The color coding indicates efficiency levels: 10\% (red), 25\% (yellow), 50\% (green), 75\% (cyan), 90\% (blue) and 100\% (magenta) in the first and second rows. For cases with $u_0>0$, the third and fourth rows show the efficiency differences relative to the $u_0=0$ cases.
	\label{fig:u0}}
\end{figure}

\subsection{Impact parameter}
\label{sec:u0}

The main analysis assumes zero impact parameter $u_0=0$, and in this section, we examine the effects of a non-zero $u_0$. 

In Figure~\ref{fig:u0}, we show \csst\ satellite detection efficiency for an Earth-mass FFP lens in the disk with a G2V source ($\rho=0.33$) and a Neptune-class disk lens with an M0V source ($\rho=0.1$), in the first and second rows of panels, respectively. The left, middle, and right sub-panels correspond to $u_0=0,\,\rho/2,\,\rho$, respectively. For cases with $u_0>0$, we present the efficiency difference relative to $u_0=0$ in the third and fourth rows. Most sensitivity zones show no significant differences, with the grazing case ($u_0=\rho$) exhibiting a distinct reduction in efficiency ($\sim40\%$) in only a small portion of the sensitivity zones. Therefore, our default analysis with $u_0=0$ effectively represents the overall cases.

\section{Existence of simulated satellites}
\label{sec:existence}

The lack of detected exomoons leaves us without evidence to determine whether planet-satellite systems in the Solar System are representative or not. This situation is analogous to the pre-1990s era of exoplanet searches, when the Solar System was the only known planetary system. Drawing from the lessons of diverse exoplanet discoveries, we adopt an approach that probes the available parameter space without making  {\it a priori}  assumptions about the distribution of exomoons. Nevertheless, the existence of detectable exomoons is subject to certain physical constraints. In this section, we examine simple physical considerations, including the Roche radius, the Hill radius, and the survival of exomoons during dynamical ejections.

The minimum orbital separation for a satellite to avoid tidal disruption is given by the planet's Roche radius, $r_{\rm Roche}=2.44 R_{\rm planet}\left({\rho_{\rm planet}}/{\rho_{\rm satellite}}\right)^{1/3}$, where $R_{\rm planet}$ is the planet's radius and $\rho_{\rm planet}$ $(\rho_{\rm satellite})$ is the density of the planet (satellite). For typical densities of planets ($\sim3\text{--}6\,{\rm g/cm^3}$ for rocky planets, $\sim0.7\text{--}1.6\,{\rm g/cm^3}$ for gas or ice giants) and satellites ($\sim1.5\text{--}3.5\,{\rm g/cm^3}$), $r_{\rm Roche}$ is several times $R_{\rm planet}$, much smaller than the minimum separations detectable planets by \csst\ or \rst. Therefore, the Roche radius constraint has no impact on our parameter space of interest.

\begin{figure}[h]
	\centering
	\includegraphics[width=1.0\textwidth]{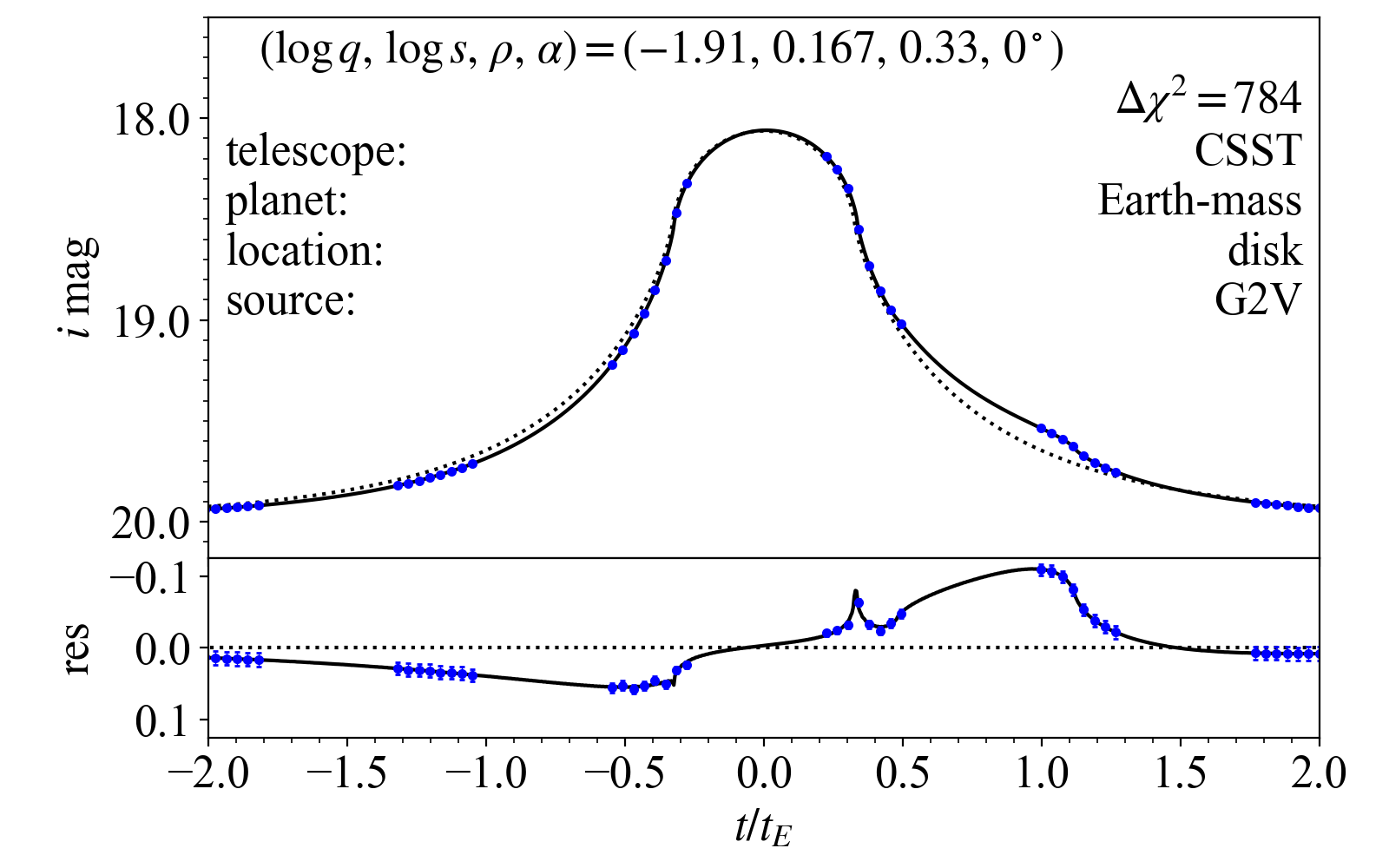}
	\caption{A mock CSST light curve of a detectable Moon-mass satellite orbiting an Earth-mass FFP in the disk. The satellite-planet separation is $r_\perp\approx0.007$\,\au, within one (one tenth) Hill radius of an Earth-mass planet at 1 (10) \,\au.}
	\label{fig:earth_moon}
\end{figure}

A satellite remains stably bound to a planet within the Hill radius: 
\begin{equation}
r_{\rm H}=a_{\rm planet}\left(\frac{m_{\rm planet}}{3m_{\rm star}}\right)^{1/3} =0.01\,{\rm AU} \left(\frac{a_{\rm planet}}{1\,{\rm AU}}\right)\left(\frac{m_{\rm planet}}{M_\oplus}\right)^{1/3}\left(\frac{m_{\rm star}}{M_\odot}\right)^{-1/3},
\end{equation}
where $a_{\rm planet}$ is planet's semi-major axis, and $m_{\rm star}$ is the mass of the host star. Dynamical simulations \citep{Debes07,Rabago19,Hong18} suggest that, the semi-major axes of planet-satellite systems generally experience only minor changes during planet-planet scattering. Thus, we assume that the ejected systems retain their original semi-major axes. 

The Einstein radii of our simulated FFPs in the disk (bulge) are
$R_\E = D_\Le \sqrt{\kappa M_\Le \pi_{\rm rel}}  = 0.007\,(0.005)\,\au\left({m_{\rm planet}}/{M_\oplus}\right)^{1/2}$.
The scaled Hill radius for the disk (bulge) cases is then given by:
\begin{equation}
  \frac{r_{\rm H}}{R_\E}=1.4\,(2.0)\left(\frac{a_{\rm planet}}{1\,{\rm AU}}\right)\left(\frac{m_{\rm planet}}{M_\oplus}\right)^{-1/6}\left(\frac{m_{\rm star}}{M_\odot}\right)^{-1/3}.
\end{equation}
Therefore, $r_{\rm H}/{R_\E}=1$ corresponds to Earth-mass planets orbiting a 1 $M_\odot$ star at $a_{\rm planet}=0.7\,(0.5)\,\au$ in the disk (bulge).
The $50\%$ satellite detection zone extends up to at most a factor of $\sim2\,(3)$ from $R_\E$ for \csst\ (\rst). Thus, for systems ejected from $a_{\rm planet} \lesssim 1\,\au$, only a minority fraction of detectable satellites in the outer part ($s\gg1$) of  the sensitivity zone are outside the Hill radius. Figure~\ref{fig:earth_moon} illustrates an example of a Moon-mass satellite detectable by \csst\ within the Hill radius of an Earth analog. For wide-separation ($\gtrsim 20\,\au$) Earth-mass planets, the entire sensitivity zone lies within the Hill radius. For Neptune-class FFPs in the disk (bulge), $r_{\rm H}/{R_\E}=1$ corresponds to $a_{\rm planet}=0.013\,(0.008)\,\au$, much smaller than typical orbital radius ($\gtrsim1\,\au$) for ejection. Therefore, the satellite detection zones for Neptune-class FFPs are entirely within the Hill radius. Note that microlensing measures the projected satellite-planet separation on the sky, and consequently, factors like inclination and eccentricity need to be taken into account in a realistic case.

Previous studies \citep{Hong18, Rabago19} have shown that some planet-satellite systems may retain their satellites during dynamical ejection. \cite{Hong18} used N-body simulations to investigate the dynamics of exomoons during planet-planet scattering, finding that most satellites with $a \lesssim 0.1 r_{\rm H}$ survive post-scattering for planets of 0.1–1 $M_{\rm J}$. Similarly, \citet{Rabago19} showed that in systems with three Jupiter-mass planets orbiting a solar-mass star, bound satellites typically have $a \lesssim 0.07\,\au$ (i.e., $\lesssim 0.1 r_{\rm H}$) after scattering. If this survival threshold of $\lesssim0.1\,r_{\rm H}$ applies to lower-mass FFPs in our simulations, most of our detectable satellites around Neptune-class planets will survive. In contrast, most satellites in the sensitivity zone for Earth-mass FFPs do not survive if ejections occur in the inner planetary system at $\sim 1$\,\au. However, dynamical considerations suggest ejections are more likely to take place in the outer planetary system, where the Safronov number exceeds unity, and most detectable satellites ejected at $\gtrsim 10$\,\au\ survive.

\section{Summary \& Discussion}
\label{sec:discussion}

In this paper, we study \csst\ detection efficiency for satellites around Earth-mass and Neptune-class FFPs, considering two representative types of G2V (solar-like) and M0V (M-dwarf) stellar sources.  For a G2V source, \csst\ is capable of detecting satellites around a Neptune-class FFP in the Galactic disk down to Moon-mass satellites near the Einstein radius, with the sensitivity zone extending over a decade in projected separation for an Earth-mass satellite. The detection efficiency in the $s-q$ plane decreases significantly for an Earth-mass FFP. Its sensitivity zones show a pronounced bi-modal shape, including some sensitivity to Moon-mass satellite at $s\sim2$. By comparison, the satellite detection efficiency reduces substantially for FFPs in the Galactic bulge or for an M0V source. \rst\ demonstrates higher sensitivity compared to \csst, particularly for M-dwarf sources, due to its infrared bandpass. The exomoon detections enabled by \csst\ and \rst\ will probe uncharted territories of exomoons around FFPs and test theoretical predictions of planetary dynamical evolutions.

In our \csst\ simulations, we assume an idealized observing strategy of continuous $i$-band exposures during a 40-min span in each orbit. After the microlensing survey strategy is finalized, more sophisticated simulations will be necessary to incorporate realistic observing patterns and the effects of multiple filters. We also note that some binary-lens perturbations could be mimicked by binary-source single-lens (1L2S) events \citep[e.g.,][]{Gaudi98}. Discussion on the possible 2L1S-1L2S degeneracy is deferred to future studies.

Finally, we briefly discuss the possibility of tidal heating. Ejection of planet-satellite systems generally results in eccentric orbits \citep{Rabago19}, which could lead to significant heating due to tidal circularization. \citet{Debes07} showed that an ejected Earth-Moon system with a semi-major axis $a\sim 0.001$\,\au\ experiences significant tidal heating --- exceeding radiogenic heating on the Earth during the first few $10^8$\,yr --- potentially making it habitable. However, such close separations are not detectable based on our simulations. Neptune-Earth systems at detectable separations $(a\sim0.01\,\au)$ could achieve similar heating rates from tidal circularization, governed by $|\dot{E}_{\rm circ}|\propto (R_{\rm planet}/a)^5 a^{-5/2} m_{\rm satellite}^2 \sqrt{m_{\rm satellite}+m_{\rm planet}}$ \citep{Debes07}. Therefore, spaced-based microlensing surveys are likely promising for detecting tidally heated Earth-mass satellites around free-floating/wide-separation icy giants, if they exist.

\normalem
\begin{acknowledgements}
We thank Andy Gould and Wei Zhu for helpful discussions and the anonymous reviewer for valuable comments. This work is supported by  the National Natural Science Foundation of China (Grant No. 12133005) and the China Manned Space Program with grant no. CMS-CSST-2025-A16. S.D. acknowledges the New Cornerstone Science Foundation through the XPLORER PRIZE.
\end{acknowledgements}
  
\bibliographystyle{raa}
\bibliography{bibtex}

\end{document}